\documentclass[aps,prb,twocolumn,showpacs,preprintnumbers,amsmath,amssymb,superscriptaddress]{revtex4-1}

\usepackage{graphicx}
\usepackage{dcolumn}
\usepackage{bm}
\usepackage{color}

\usepackage{ulem}
\renewcommand{\emph}[1]{\textit{#1}}

\definecolor{darkblue}{rgb}{0,0,0.5}
\definecolor{darkgreen}{rgb}{0,0.5,0}
\definecolor{darkred}{rgb}{.7,0,0}
\definecolor{purple}{rgb}{0.5,0,0.6}
\definecolor{orange}{rgb}{1,0.5,0}
\definecolor{grey}{rgb}{.6,.6,.6}
\definecolor{lightpink}{rgb}{1,0.7,0.75}
\definecolor{pink}{rgb}{1,0.4,0.58}
\definecolor{deeppink}{rgb}{1,0.08,0.58}

\newcommand{\bra}[1]{\langle #1|}
\newcommand{\ket}[1]{|#1\rangle}
\newcommand{\bracket}[2]{\langle #1 \vert #2 \rangle}

\newcommand{\expect}[1]{\langle #1 \rangle}

\newcommand{\be}{\begin{equation}}
\newcommand{\ee}{\end{equation}}
\newcommand{\bea}{\begin{eqnarray}}
\newcommand{\eea}{\end{eqnarray}}

\newcommand{\eq}[1]{Eq.~(\ref{#1})}
\newcommand{\app}[1]{App.~\ref{#1}}

\newcommand{\Fig}[1]{Figure~\ref{#1}}
\newcommand{\fig}[1]{Fig.~\ref{#1}}
\newcommand{\figs}[1]{Figs.~\ref{#1}}

\newcommand{\Sec}[1]{Sec.~\ref{#1}}

\newcommand{\e}{\varepsilon}
\newcommand{\w}{\omega}
\newcommand{\s}{\sigma}
\newcommand{\up}{\uparrow}
\newcommand{\down}{\downarrow}

  \definecolor{cwc}{rgb}{.3,.9,.3}
  \definecolor{cwb}{rgb}{.8,.4,.1}

\RequirePackage[
   hyperindex,colorlinks,bookmarksnumbered,
   plainpages=true,pdfstartview=FitH]{hyperref}
\hypersetup{linkcolor=blue,urlcolor=blue,citecolor=blue}
\usepackage{hyperref}

\def\nzero{n_0}
\newcommand{\trace}{\mathop{\mathrm{Tr}}\nolimits}
\newcommand{\ie}{i.\,e. }

\newcommand{\wrt}{w.r.t. }
\newcommand{\vs}{vs. }
\newcommand{\cf}{cf. }
\newcommand{\Hloc}{\hat{H}_\mathrm{loc}}

\newcommand{\TK}{T_{K}}

\begin{document}

\title{Thermalization and dynamics in the single impurity Anderson model}

\author{Ireneusz Weymann}
\email{weymann@amu.edu.pl}
\affiliation{Physics Department, Arnold Sommerfeld
Center for Theoretical Physics and Center for NanoScience, \\
Ludwig-Maximilians-Universit\"at, Theresienstrasse 37, 80333
Munich, Germany}
 \affiliation{Faculty of Physics, Adam Mickiewicz University, 61-614 Pozna\'n, Poland}

\author{Jan von Delft}
\affiliation{Physics Department, Arnold Sommerfeld
Center for Theoretical Physics and Center for NanoScience, \\
Ludwig-Maximilians-Universit\"at, Theresienstrasse 37, 80333
Munich, Germany}

\author{Andreas Weichselbaum}
\affiliation{Physics Department, Arnold Sommerfeld
Center for Theoretical Physics and Center for NanoScience, \\
Ludwig-Maximilians-Universit\"at, Theresienstrasse 37, 80333
Munich, Germany}

\date{\today}

\begin{abstract}
  We analyze the process of thermalization, dynamics and the
  eigenstate thermalization hypothesis (ETH) for the single
  impurity Anderson model, focusing on the Kondo regime.  For this
  we construct the complete eigenbasis of the Hamiltonian using
  the numerical renormalization group (NRG) method in the language
  of the matrix product states. It is a peculiarity of the NRG
  that while the Wilson chain is supposed to describe a
  macroscopic bath, very few single particle excitations already
  suffice to essentially thermalize the impurity system at finite
  temperature, which amounts to having added a macroscopic
  amount of energy. Thus
  given an initial state of the system such as the ground state
  together with microscopic excitations, we calculate the spectral
  function of the quantum impurity using the microcanonical and
  diagonal ensembles. These spectral functions are compared to the
  time-averaged spectral function obtained by time-evolving the
  initial state according to the full Hamiltonian, and to the
  spectral function calculated using the thermal density matrix.
  By adding or removing particles at a certain Wilson
  energy shell on top of the ground state,
  we find qualitative
  agreement between the resulting spectral functions
  calculated for different ensembles. This
  indicates that the system thermalizes
  in the long-time limit, and can be
  described by an appropriate statistical-mechanical ensemble.
  Moreover, by calculating static quantities such as the impurity
  spectral density at the Fermi level as well as the dot occupancy
  for energy eigenstates relevant for microcanonical ensemble, we
  find good support for ETH. The ultimate mechanism 
  responsible for this effective thermalization within
  the NRG can be identified as Anderson orthogonality:
  the more charge that needs to flow to or from infinity
  after applying a local excitation within the Wilson chain,
  the more the system looks
  thermal afterwards at an increased temperature. For the same
  reason, however, thermalization fails if charge rearrangement
  after the excitation remains mostly local. In these cases,
  the different statistical ensembles lead to different results.
  Their behavior needs to be understood as
  a plain microscopic quantum quench only.
\end{abstract}



\maketitle

\section{Introduction}

In an isolated many-body quantum system, the dynamics can drive the
system towards a stationary state that resembles a thermal state.
\cite{sengupta04,berges04} Its resulting macroscopic properties are
therefore consistent with statistical mechanics. \cite{krylov,huang}
Thermalization exhibits a certain universality, in that widely
different initial conditions can lead to very similar thermal
states. \cite{Deutsch,Srednicki,Rigol_Nature08} Several recent
studies have aimed at finding the reason for this universality.
\cite{Rigol_Nature08,rigol09,cassidy11, rigol12, neuenhahn12,
riera12, ziraldo13} It has been argued that for generic isolated
interacting quantum systems the process of thermalization
is described by eigenstate thermalization hypothesis (ETH).\cite{Deutsch,Srednicki}
ETH states that expectation values of generic observables calculated
using a {\it diagonal} ensemble derived from a single generic
many-body state and characterized by a density matrix that is
diagonal in energy eigenbasis, and those calculated for a proper
energy eigenstate, are equal. This happens if eigenstates that are
close in energy yield similar expectation values for the operator in
question. 
It implies that the knowledge of a single eigenstate with energy falling within
a proper energy window is sufficient to describe the system in its
thermal state.
Although for systems displaying {\it simple} correlations
the ETH may be rather intuitive,~\cite{Rigol_Nature08} it is
definitely not obvious for fully-interacting many-body problems, the
description of which involves exponentially large Hilbert spaces.

In this paper we study the process of thermalization and explore the
applicability of the ETH for a particular interacting many-body
system, namely the single-impurity Anderson model (SIAM).
For this model, a complete basis of approximate eigenstates
of the Hamiltonian can be constructed using the
numerical renormalization group (NRG) method.
\cite{WilsonRMP75,BullaRMP08}
The availability of
this complete basis of energy eigenstates allows numerous aspects
of the ETH to be analyzed in great detail.

The SIAM describes a localized electronic level (the {\it quantum impurity},
henceforth called quantum dot) with local interactions, that
hybridizes with a band of conduction electrons.  Proposed originally
to explain the formation of local moments in metals,~\cite{Anderson}
this model and generalizations thereof have been commonly used to
model transport through small quantum dots coupled to leads.
\cite{krishnamurthy80,meir,meir93} When the local level, henceforth
called dot level, has average occupancy $n_{d\sigma} = \expect{
\hat{d}_\s^\dag \hat{d}_\s} \simeq \frac{1}{2}$ per spin species, it
hosts a localized spin which experiences exchange interactions with
the spins of the conduction band, leading to the Kondo
effect.~\cite{kondo64,hewson_book} At temperatures below the Kondo
temperature, $T_K$, the conduction electrons screen the dot's spin,
forming a nonlocal singlet state.~\cite{kondo64,hewson_book} The
many-body correlations involved in this screening can be
characterized~\cite{hewson_book} in terms of the local density of
states (LDOS). For the ground state $| \mathcal{G} \rangle$, the
LDOS is given by the local spectral function, defined as the
expectation value $ A_\mathcal{G} = \langle \mathcal{G} |
\hat{\mathcal{A}} | \mathcal{G} \rangle$ of the spectral operator
$\hat{\mathcal{A}} \equiv \sum_\sigma ( \hat{\mathcal{A}}_\sigma^{+}
+ \hat{\mathcal{A}}_\sigma^{-} )$, where
\begin{eqnarray}  \label{Eq:Aoperator}
  \hat{\mathcal{A}}_\sigma^{+}
  &=& \hat{d}_\sigma \,\delta(\omega - \hat{H} +E_0)\,
      \hat{d}_\sigma^\dag
\notag\\
  \hat{\mathcal{A}}_\sigma^{-}
  &=& \hat{d}_\sigma^\dag \,\delta(\omega + \hat{H} -E_0 )\,
      \hat{d}_\sigma\text{,}
\end{eqnarray}
with $\delta(\w)$ the Dirac delta function, $\hat{H}$ the
Hamiltonian, and $E_0$ the ground state energy. The spectral
operator $\hat{\mathcal{A}} $ measures the likelihood of raising or
lowering the energy by $\w$ upon adding or removing an electron from
the dot. When $A_\mathcal{G}$ is viewed as function of $\w$
(implicit in our notation) the ground state spectral function
exhibits a striking, sharp peak near the Fermi level, the so-called
Kondo-Abrikosov-Suhl resonance,~\cite{hewson_book} whose width is a
measure of the Kondo temperature, $T_K$.  This resonance is
characteristic for the spectral function
\begin{eqnarray}
 A_{\Psi} = \langle \Psi | \hat {\mathcal{A} } | \Psi \rangle
\end{eqnarray}
of any quantum state $|\Psi \rangle$ for which the average energy
and energy uncertainty,
\begin{subequations}\label{eq:EPsi}
\begin{eqnarray}
   E_\Psi &=& \bra{\Psi} \hat{H} -E_0 \ket{\Psi} \\
   \Delta E_\Psi &=&
   \sqrt{ \bra{\Psi} (\hat{H} -E_0)^2 \ket{\Psi} - E_\Psi^2}
\end{eqnarray}
\end{subequations}
both lie below the Kondo temperature (meaning the energy of the
system when $T=\TK$).
However, the resonance weakens and eventually disappears once
$E_\Psi$ and/or $\Delta E_\Psi$ becomes larger than $\TK$ (examples
are shown below). As a result, the Kondo resonance causes a striking
enhancement in the transmission through the dot if the temperature
and source-drain bias are lowered to become smaller than $T_K$,
causing a zero-bias anomaly that has been observed in numerous
experiments.~\cite{goldhaber-gordon_98,cronenwett_98,derWiel_Science00}

An accurate description of Kondo correlations in general and the
LDOS in particular requires sophisticated theoretical tools.
In this regard, the NRG 
has proven itself to be a particularly powerful and versatile
method for studying various quantum impurity models.
\cite{WilsonRMP75,BullaRMP08} 
NRG employs a logarithmic discretization of the conduction
band and maps the model onto a tight-binding chain with
exponentially decaying hoppings, $t_n \sim \Lambda^{-n/2}$ with
dimensionless discretization parameter $\Lambda \gtrsim 2$, the
so-called Wilson chain.  This chain is diagonalized iteratively by
adding one site at a time,~\cite{WilsonRMP75} and the eigenstates
calculated during this iterative scheme~\cite{anders05a,anders05b}
can be used to construct a complete many-body basis of approximate
eigenstates of the Hamiltonian, the Anders-Schiller (AS) basis.
\cite{anders05a,anders05b} Using the AS-basis, it is possible to
accurately calculate spectral functions of local operators
\cite{WeichselbaumPRL07,petersPRB06} and in particular to reliably
determine the shape of the Kondo resonance.  Moreover, the time
evolution $\ket{\Psi_t}$ of an arbitrary initial quantum state
$\ket{\Psi}$ can be calculated by representing the latter in the
AS-basis.~\cite{anders05a,anders05b} With this approach, called
time-dependent NRG (tNRG), it is possible to explicitly calculate
the evolution of operator expectation values, such as that of the
spectral operator $\hat{\mathcal{A}}$.

Since tNRG treats the entire many-body Hilbert space as a closed
quantum system when calculating the time-evolution $\ket{\Psi_t}$,
it is ideally suited to studying thermalization and the ETH for
interacting many-body systems. In this paper, we do this for the
SIAM. We consider a variety of initial states $\ket{\Psi}$, some
correlated, some not, explore how Kondo correlations emerge in the
long-time limit, and analyze to what extent the results correspond
to those expected for a thermal state. We will use the
time-dependent expectation value of the spectral operator,
\begin{equation}
 A_t = \bra{\Psi_t} \hat{\mathcal{A}} \ket{\Psi_t} \; ,
\end{equation}
as diagnostic tool for the emergence of Kondo correlations with
time: they lead to the emergence of a Kondo resonance in $A_t$, when
viewed as function of $\w$ for a series of fixed but ever larger
values of $t$. Moreover, the width of the Kondo resonance in $A_t$
for $t\to\infty$ can serve as a measure of the effective temperature
of the system in the long-time limit as long as this width is larger
than $\TK$. Note, however, that the simultaneous $t$ and
$\w$-dependence of $A_t$ will presumably not be accessible
experimentally; thus, $A_t$ is to be regarded mainly as a useful
diagnostic tool for theoretical analysis.

To study the dynamics of the system and assess the applicability of
ETH, we compare the long-time limit of the LDOS to the LDOS
calculated within three different ensembles: (i) diagonal ensemble
which is characterized by a density matrix that is diagonal in the
energy eigenbasis; (ii) the microcanonical ensemble corresponding to
a fixed energy, $E_{\Psi}$, cf. \eq{eq:EPsi}; and (iii) the standard
thermal grand canonical ensemble\cite{WeichselbaumPRL07} at
comparable temperature $T\sim E_{\Psi}$.
We study the process of
thermalization for the ground state and a few excited states, where
the excitations are created either in the bath or at the impurity.
More specifically, we consider the states which are created by
acting with a single-particle operator
or a density operator
on the full many-body ground state of the system. We also analyze the dynamics
and long-time behavior of the system for an initial state in which
the dot is decoupled from the leads, and for a state with
single-particle excitation in the dot. We show that when starting
the time evolution with a state for which the excitation was created
within the Wilson chain by adding or removing charge,
we get a local density of states that is similar to that obtained with a
microcanonical ensemble of corresponding energy. By calculating the
expectation values of the spectral function at the Fermi level and
the dot occupancy for energy eigenstates relevant for the
microcanonical ensemble, we demonstrate that, indeed, 
thermalization occurs and that ETH applies.
However, a rather different behavior is observed for 
states that only involve local rearrangement
of any of the conserved charges.
We show that here
the time evolution needs to be interpreted as a
microscopic quantum quench that occurs on top of a given initial
statistical ensemble.

The paper is organized as follows. In \Sec{sec:theory} we describe
the general concept of the ETH, the model and Hamiltonian, as well
as the different ensembles used to study the dynamics of the system.
Section \ref{sec:results} is devoted to testing the process of
thermalization and the ETH for different initial states. The
conclusions are given in \Sec{sec:conclusions}, while the details of
the construction of respective statistical ensembles using the NRG
eigenbasis are presented in the Appendix.

\section{Theoretical description \label{sec:theory}}

\subsection{General statement of ETH}

Consider a generic isolated quantum system with Hamiltonian
$\hat{H}$, with its complete many-body eigenbasis given by
$\ket{s}$, \ie $\hat{H} \ket{s} = E_s\ket{s}$. Suppose the system is
initialized at time $t = 0$ in a pure quantum state, $\ket{\Psi} =
\sum_s C_s \ket{s}$, with $C_s = \langle s|\psi\rangle$. To what
extend does the expectation value of an observable
$\hat{\mathcal{O}}$,
\begin{equation}
  O_t = \bra{\Psi_t} \hat{\mathcal{O}} \ket{\Psi_t} \,,
\end{equation}
with $\ket{\Psi_t} \equiv e^{-i\hat{H} t} \ket{\Psi}$, depend on the
initial state $\ket{\Psi}$ in the long-time limit, $t\to\infty$ ?

Since the system is closed, quantum mechanics gives the answer
\begin{equation} \label{Oinfty}
  O_\infty = \sum_s |C_s|^2 \mathcal{O}_{ss} \,,
\end{equation}
where $\mathcal{O}_{ss'} \equiv \bra{s} \hat{\mathcal{O}} \ket{s'}$,
which assumes that the contribution of the off-diagonal matrix
elements averages to zero due to phase cancellations, \ie
$\lim_{t\to\infty} \sum_{s \neq s'} e^{i(E_s - E_{s'} )t}
\mathcal{O}_{ss'} = 0$.
Thus, the long-time limit is described by a diagonal ensemble,
characterized by a density matrix that is diagonal in the energy
eigenbasis,
\begin{subequations}\label{Odiag}
\begin{eqnarray}
  O_{\rm diag} &=& {\rm Tr} [ \hat{\rho}_{\rm diag} \hat{\mathcal{O}} ]
\\[1ex] \text{with} \quad
  \hat{\rho}_{\rm diag} &\equiv& \sum_s |C_s|^2  \hat{P}_s \,,
\end{eqnarray}
\end{subequations}
where $\hat{P}_s \equiv \ket{s}\bra{s}$ is the projector onto
eigenstate $\ket{s}$.

However, standard statistical mechanics gives a different answer: it
assumes that the long-time limit is well described by a
microcanonical ensemble,
\begin{subequations} \label{Omicro}
\begin{eqnarray}
  O_{\rm micro} &=& {\rm Tr}
  [ \hat{\rho}_{\rm micro} \hat{\mathcal{O}} ]
\\[1ex] \text{with} \quad
 \hat{\rho}_{\rm micro} &\equiv&
 \tfrac{1}{N_\Psi} 
 \sum_{|E_s - E_\Psi| \leq
 \delta E_\Psi
 }
\hat P_s \,,
\end{eqnarray}
\end{subequations}
characterized by the initial state's average energy $E_\Psi$ and a
(narrow) energy window $\delta  E_\Psi \ll \Delta E_\Psi$ [\cf
\eq{eq:EPsi}], with $N_\Psi$ the number of states in the energy
window that is being summed over. The system is said to behave
thermally if in the long-time limit the expectation values
calculated using diagonal and microcanonical ensembles are equal,
\ie $O_{\rm diag} = O_{\rm micro}$. One possible scenario for which
this occurs, discussed by Deutsch and Srednicki, \cite{Deutsch,
Srednicki} is the eigenstate thermalization hypothesis, which
assumes that the eigenstate expectation values $\mathcal{O}_{s s}$
show only very weak state-to-state fluctuations for states that are
close in energy. If this is the case, the knowledge of the
expectation value $\mathcal{O}_{s s}$ of observable $\mathcal{O}$
for a \emph{single} eigenstate with appropriately chosen energy is
sufficient to describe the system's behavior in the long time limit.

Given the difference in form between Eqs.~(\ref{Odiag}) and
(\ref{Omicro}), the intriguing question arises: under what
conditions are the two equivalent? Does the eigenstate
thermalization hypothesis also hold for truly interacting,
nontrivial many-body systems? The main goal of this paper is to
address these questions for the single impurity Anderson
Hamiltonian.

\subsection{Model Hamiltonian and NRG basics}

\subsubsection{Model Hamiltonian}

The SIAM describes a single-level quantum dot that is tunnel-coupled
to a conduction band of free electrons. Its Hamiltonian takes the
form~\cite{Anderson}
\begin{subequations} \label{Eq:HAnderson}
\bea \label{Eq:HAnderson:all}
\hat{H} &=&  \hat{H}_{\rm lead} +
   \underset{ \equiv \Hloc }{\underbrace{
      \hat{H}_{\rm dot} + \hat{H}_{\rm tun} }
   } \,,
\eea
where
\bea
  \hat{H}_{\rm lead} &=&
      \sum_{k\s} \e_{k} \hat{c}^\dag_{k\s} \hat{c}_{k\s}
      \label{Eq:H:lead}
\\
  \hat{H}_{\rm dot} &=&
      \sum_{\s} \e_{d} \hat{d}^\dag_\s \hat{d}_\s
    + U \hat{d}^\dag_\up \hat{d}_\up \hat{d}^\dag_\down \hat{d}_\down
      \label{Eq:H:Dot}
\\
  \hat{H}_{\rm tun} &=&
      \sum_{k\s} v \left(\hat{c}^\dag_{k\s} \hat{d}_\s
    + \hat{d}^\dag_\s \hat{c}_{k\s} \right)
      \label{Eq:H:Tun}
  \text{.}
\eea
\end{subequations}
$\hat{H}_{\rm lead}$ describes a band of noninteracting conduction
electrons, where $\hat{c}^\dag_{k\s}$ creates an electron with spin
$\sigma\in\{\uparrow,\downarrow\}$ at energy $\e_{k}$. $\hat{H}_{\rm
dot}$ describes a dot state with level energy $\e_{d}$, creation
operator $\hat{d}^\dag_\s$ and local Coulomb energy $U$.
$\hat{H}_{\rm tun}$ accounts for tunneling between the dot and the
conduction band, with energy-independent hopping matrix element $v$.
Therefore $\Hloc$ accounts for the local part of the Hamiltonian
that includes the quantum dot as well as the bath degrees of freedom
at the location of the impurity.

If $n_{d\sigma} \simeq \frac{1}{2}$, the electronic correlations can
lead to the Kondo effect at low temperatures, when due to spin-flip
cotunneling processes the dot's spin becomes screened by the
conduction electrons. To resolve the Kondo physics, we use NRG.
\cite{WilsonRMP75,BullaRMP08,anders05a,anders05b,WeichselbaumPRL07}
There the total system Hamiltonian is described by $\hat{H} \equiv
\lim_{N\to\infty}\hat{H}_N$ with
\begin{subequations} \label{Eq:HNRG}
\be \label{Eq:HNRG:N}
  \hat{H}_{N} = \Hloc
  + \sum_{n=0}^{N-1} t_n \sum_\s  \left( \hat{f}_{n\s}^\dag \hat{f}_{n+1\s}
  + \mathrm{H.c.} \right),
\ee
such that $\hat{H}_0 \equiv \Hloc$, where
\be \label{Eq:HNRG:Loc}
  \Hloc = \hat{H}_{\rm dot}
  + \sqrt{\tfrac{2 \Gamma}{\pi}} \, \sum_\s
    \left( \hat{d}_{\s}^\dag \hat{f}_{0\s} + \mathrm{H.c.} \right) \,.
\ee
\end{subequations}%
The operators $\hat{f}_{n\s}$ act on the $n$th shell of the Wilson
chain, with $t_n$ the respective nearest-neighbor hoppings along the
chain. For simplicity, a constant hybridization strength $\Gamma
(\omega) \equiv \Gamma \vartheta(D-|\omega|)$ is assumed, with level
half-width $\Gamma \equiv \pi\rho|v|^2 $ and half-bandwidth $D:=1$,
taken as the unit of energy throughout. Similarly, units of time and
temperature are fixed by using $\hbar = k_B=1$, respectively.

\subsubsection{Complete basis sets}

\begin{figure}[t]
\includegraphics[width=0.8\columnwidth]{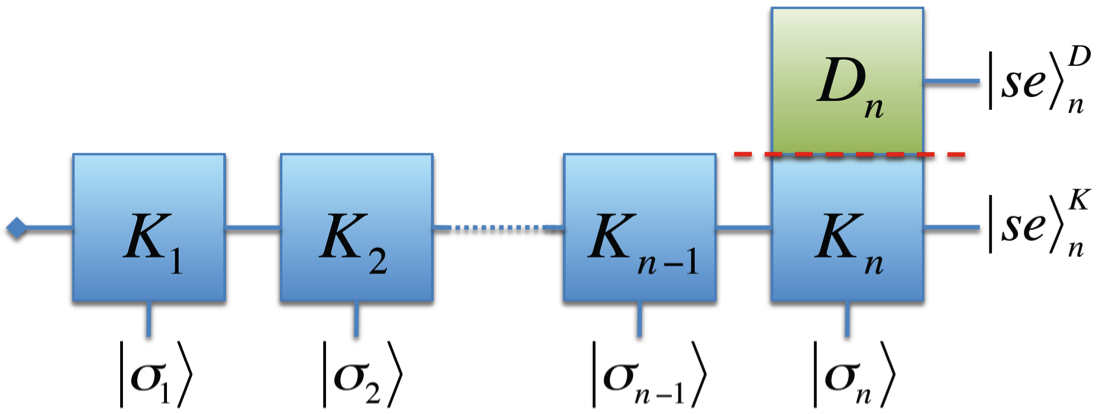} \caption{
  \label{Fig:mps} (color online) Matrix product state (MPS)
  representation of the kept (K) and discarded (D) states on the
  Wilson chain. The $n$th box represents the tensor block $X_n$ ($X
  \in\{K,D\}$), while its bottom, left and right legs carry the labels
  of the local states $\ket{\sigma_n}$, the kept states
  $\ket{se}_{n-1}^K$, and the kept (discarded) states $\ket{se}_{n}^K$
  ($\ket{se}_{n}^D$), respectively.}
\end{figure}

The NRG Hamiltonian in \eq{Eq:HNRG} can be solved in an iterative
way, by constructing and diagonalizing the Hamiltonian shell-wise.
Due to exponentially growing Hilbert space with increasing the
iteration number $n$, one needs to truncate the space by keeping
only $M_K$ lowest-energy states at a given iteration. Thus, each
Wilson shell, up to the last iteration, $n=N$, contains the
low-energy states retained for the next iteration, \ie the kept ($K$)
states, and the discarded ($D$) states (starting from shells, $n\geq
\nzero$, where $\nzero$ is the iteration at which one first begins
to truncate), while at the last shell (for $n=N$) all states are
regarded as discarded.  The discarded states can be used to define a
complete eigenbasis of the Hamiltonian,
\begin{subequations} \label{Eq:closrel}
\begin{equation}
  {\bf 1} = \sum_{n \geq \nzero}^N \sum_{se} \ket{se}_n^{D} {}_n^{D} \bra{se},
\end{equation}
which, to a good approximation due to presence of energy scale
separation along the Wilson chain, also forms an eigenbasis of the
full Hamiltonian, \cite{anders05a,anders05b}
\begin{equation}
  \hat{H}\ket{se}_n^{X} \simeq E_{ns}^{X} \ket{se}_n^{X}.
\label{eq:NRG:approx}
\end{equation}
\end{subequations}
Here $\ket{se}_n^{X}$ denotes a kept or discarded ($X\in{K,D}$)
state $s$ on the Wilson chain at shell $n$, $e$ is an index labeling
environmental states describing the rest of the chain (shells with
larger $n$), and $E_{ns}^{X}$ is the corresponding approximate
eigenenergy.  Note that the discarded states are defined on the
Wilson shells from $n=\nzero$ to $n=N$.  The states obtained by
diagonalizing the NRG Hamiltonian can be conveniently expressed in
terms of matrix product states
(MPS),~\cite{WeichselbaumPRL07,WeichselbaumPRB09} see
Fig.~\ref{Fig:mps}.
The coefficients of any state $|\Psi\rangle$ in this
complete eigenbasis will be denoted by $C_{nse}^{D}
\equiv {}^{D}_n \bracket{se}{\Psi}$.

\subsubsection{Model parameters}

We use the following parameters throughout this paper: $U=0.12 $,
$\Gamma=0.01 $, and $\e_d=-U/3$. This yields the Kondo
temperature,~\cite{Haldane1978} $T_K \simeq 3.7 \times 10^{-4}$ (in
units of $D=1$). For the Wilson chain we take the following
parameters: the discretization $\Lambda=2$, the chain length $N=50$,
and we keep $M_K=1024$ states for each iteration, unless stated
otherwise. Moreover, in calculations we used the Abelian symmetry
for the total charge and the total spin $z$-th component. Using the
complete eigenbasis, we calculate the relevant matrix elements and
collect the discrete spectral function data points in a dense
logarithmic mesh. In order to represent the continuum model, this
discrete data subsequently needs to be broadened. Here we follow the
standard prescription \cite{BullaRMP08,WeichselbaumPRL07} of
approximating the Dirac delta function by log-Gaussian
distributions, while smoothly switching to a standard Gaussian
distribution for $|\w|<\w_0$ in order to broaden across $\omega=0$,
with $\w_0 = E_\Psi$, unless stated otherwise. The latter is
required since the resolution of NRG below temperature -- here more
broadly the energy of the system -- is intrinsically limited by
energy scale separation. \cite{WeichselbaumPRL07}

\subsection{Density matrices}\label{Sec:DM}

Using the complete eigenbasis of the NRG Hamiltonian, it is possible
to calculate the thermal grand canonical expectation value $O_{\rm
grand} = {\rm Tr} [\hat \rho_{\rm grand} \hat {\mathcal{O}}]$ of any
operator, with the grand canonical density matrix represented
as~\cite{WeichselbaumPRL07}
\begin{eqnarray} \label{Eq:rhogrand}
  \hat{\rho}_{\rm grand} &=& \sum_{n}\sum_{se}   \ket{se}_n^{D} \;
  \frac{e^{-\beta E_{ns}^{D}}}{Z} \; {}_n^{D} \bra{se}\nonumber\\
  &\equiv& \sum_{n} w^{\rm grand}_n \hat{\rho}^{\rm grand}_{n}\;,
\end{eqnarray}
with the chemical potential set to zero, $\mu=0$, and $\beta = 1/T$. Here
$\hat{\rho}^{\rm grand}_{n}$ is the density matrix of shell $n$
within the state space $s\in D$, $w^{\rm grand}_n = d^{N-n} Z_n / Z$
is the total weight of shell $n$, with the partition function $Z_n$
and $d=4$ the dimension of the local state space of a Wilson site.
Henceforth, we use the abbreviated notation $\sum_n \equiv
\sum_{n\geq \nzero}^N$ for the sum over the Wilson shells with
$n\geq \nzero$. This notation will be used for all ensembles
considered in this paper.

With the complete eigenbasis, the density matrices for the
microcanonical and diagonal ensembles can also be constructed.
Similarly to $\hat{\rho}_{\rm grand}$, any of the ensembles
$\mathcal{E}$ has a decomposition over the Wilson shells of the
general form
\begin{equation}
    \hat{\rho}_{\mathcal{E}}
  = \sum_n w_n^{\mathcal{E}} \, \hat{\rho}^{\mathcal{E}}_n \,.
\label{Eq:rho}
\end{equation}
The NRG specific details of the implementation of the aforementioned
ensembles can be found in \app{App:AdiagAmicro}.

\section{Results and discussion \label{sec:results}}

In the following we study the dynamics of the system
for various excitations on top of the ground state of
the system. In particular, this includes
single-particle excitation by adding or removing a particle using
$\ket{\Psi} = \hat{f}_{n\s}^\dag \ket{\mathcal{G}}$
(\Sec{sec:x:1particle}), density-like excitations such as
 $\ket{\Psi} = \hat{f}_{n\s}^\dag\hat{f}_{n\s} \ket{\mathcal{G}}$
(\Sec{sec:x:density}),
and quantum quenches in the hybridization of the impurity
(\Sec{sec:x:quench}). Here the order
in which we discuss these cases is chosen such that 
displaced charge, \ie charge that needs to flow to or from
infinity, is maximal in the first case while it
is smallest in the last case. Accordingly as will be
shown below, thermalization and hence ETH is best
satisfied in the first case, but not in the last.

The expectation value we are interested in is the full many-body
spectral function $A_{\mathcal{E}}(\omega) $
of the dot level (local density of
states, LDOS) for a specified ensemble $\mathcal{E}$,
as this is most sensitive to 
correlations within the system.
Here $A(\omega) = -\frac{1}{\pi} \sum_\s \textrm{Im} \, G^R_{d\s}(
\omega)$, where $G^R_{d\s}(\omega)$ denotes the Fourier transform of
the retarded Green's function, $G^R_{d\s}(\tau) = -i\theta(\tau)
\expect{ \{\hat{d}_\sigma^{}(\tau), \hat{d}_\sigma^\dag(0)\}}_\mathcal{E}$. 
The spectral operator $\hat{\mathcal{A}}$ is 
defined by 
\be
  \hat{\mathcal{A}} = \frac{1}{\pi} \sum_\s \int_0^\infty \!\! \! d\tau \,
  \{\hat{d}_\sigma^{}(\tau),\hat{d}_\sigma^\dag(0)\} \, e^{i\w\tau} \,.
\ee
which for the ground state of the system yields \eq{Eq:Aoperator}.
The corresponding spectral function is given by
\begin{equation}
  A_{\mathcal{E}}
= {\rm Re} \{ {\rm Tr} [\hat{\rho}_{\mathcal{E}} \, \hat{\mathcal{A}}] \}\,,
\end{equation}
with the density matrix $\hat{\rho}_{\mathcal{E}}$ for a specific
statistical ensemble $\mathcal{E}$ given by \eq{Eq:rho}, or for a
single state $\ket{\Psi_t}$ by $\hat{\rho}_{\Psi_t} = \ket{\Psi_t}
\bra{\Psi_t}$. 

Here three different ensembles have been analyzed:
the grand canonical, the microcanonical and the diagonal ensemble.
The spectral function for the grand canonical ensemble, $A_{\rm grand}$,
was calculated at an effective temperature $T_{\Psi}$, such that $E_\Psi
= {\rm Tr}[\hat{\rho}_{\rm grand} \hat{H}]$, unless stated
otherwise. As in \eq{eq:EPsi}, the energy $E_\Psi$
was measured relative to the ground state energy $E_0$ throughout.
The calculation of the spectral function for the diagonal and
microcanonical ensembles using the respective density matrices can
be preformed in a way similar to the {\it full} density matrix
calculations,~\cite{WeichselbaumPRL07} see \app{App:AdiagAmicro}.
Having generated the eigenbasis of the NRG Hamiltonian in a forward
sweep over the Wilson chain, a subsequent backward sweep is
performed to determine the respective density matrices with
corresponding weights. This basically enables the calculation of all
relevant operator expectation values including, in particular, the
expectation value of the spectral function operator
$\hat{\mathcal{A}}$.

Moreover, we have also calculated the spectral function for the
initial state, $A_\Psi = \bra{\Psi} \mathcal{\hat{A}} \ket{\Psi}$,
and for the time-evolved state in the long time limit,
\be \label{Eq:Atime}
   A_{\rm time} = \frac{1}{\delta t}
   \int_{t_{\rm fin}-\delta t}^{t_{\rm fin}} dt\; A_{t},
\ee
where $t_{\rm fin}$ is the final time and $\delta t$ is the time
over which the averaging is performed. In calculations we use
$t_{\rm fin} T_K = 10^4$ and $\delta t = 0.2 t_{\rm fin}$. The
details of the calculation of time-averaged spectral function are
presented in \app{Ap:At}.

\begin{figure}[t]
\includegraphics[width=0.9\columnwidth]{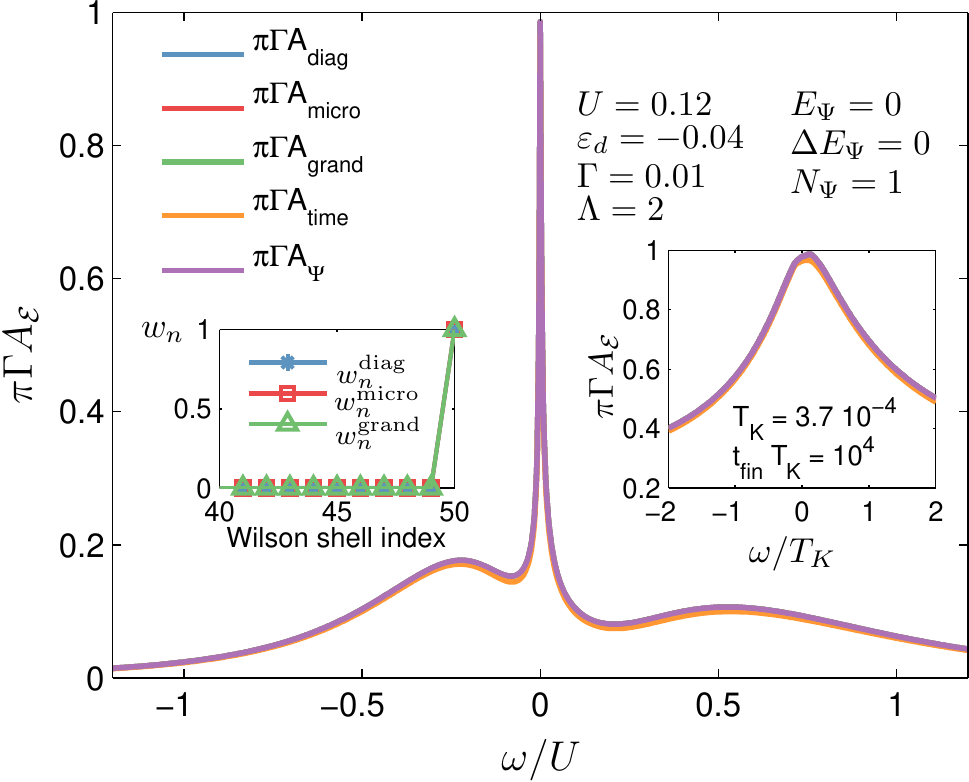}
  \caption{\label{Fig:A_for_GS} (color online) The normalized spectral
  function
  $\pi\Gamma A_{\mathcal{E}}$ 
  of the dot level at $T=0$ calculated by
  using the diagonal ensemble $A_{\rm diag}$, the microcanonical
  ensemble $A_{\rm micro}$, the grand canonical ensemble $A_{\rm
  grand}$, the time-evolved state in the long time limit $A_{\rm
  time}$, as well as the spectral function $A_\Psi$ calculated
  for the initial state $\ket{\Psi} = \ket{\mathcal{G}}$. The left inset
  displays the weights on the Wilson chain $w_n$ for the diagonal
  ($w_n^{\rm diag}$), microcanonical ($w_n^{\rm micro}$) and grand
  canonical $(w_n^{\rm grand})$ ensembles for fixed finite length
  $N=50$, while the right inset represents a plain zoom of the
  spectral data around the Kondo peak.
}
\end{figure}

\begin{figure*}
\includegraphics[width=2.0\columnwidth,clip]{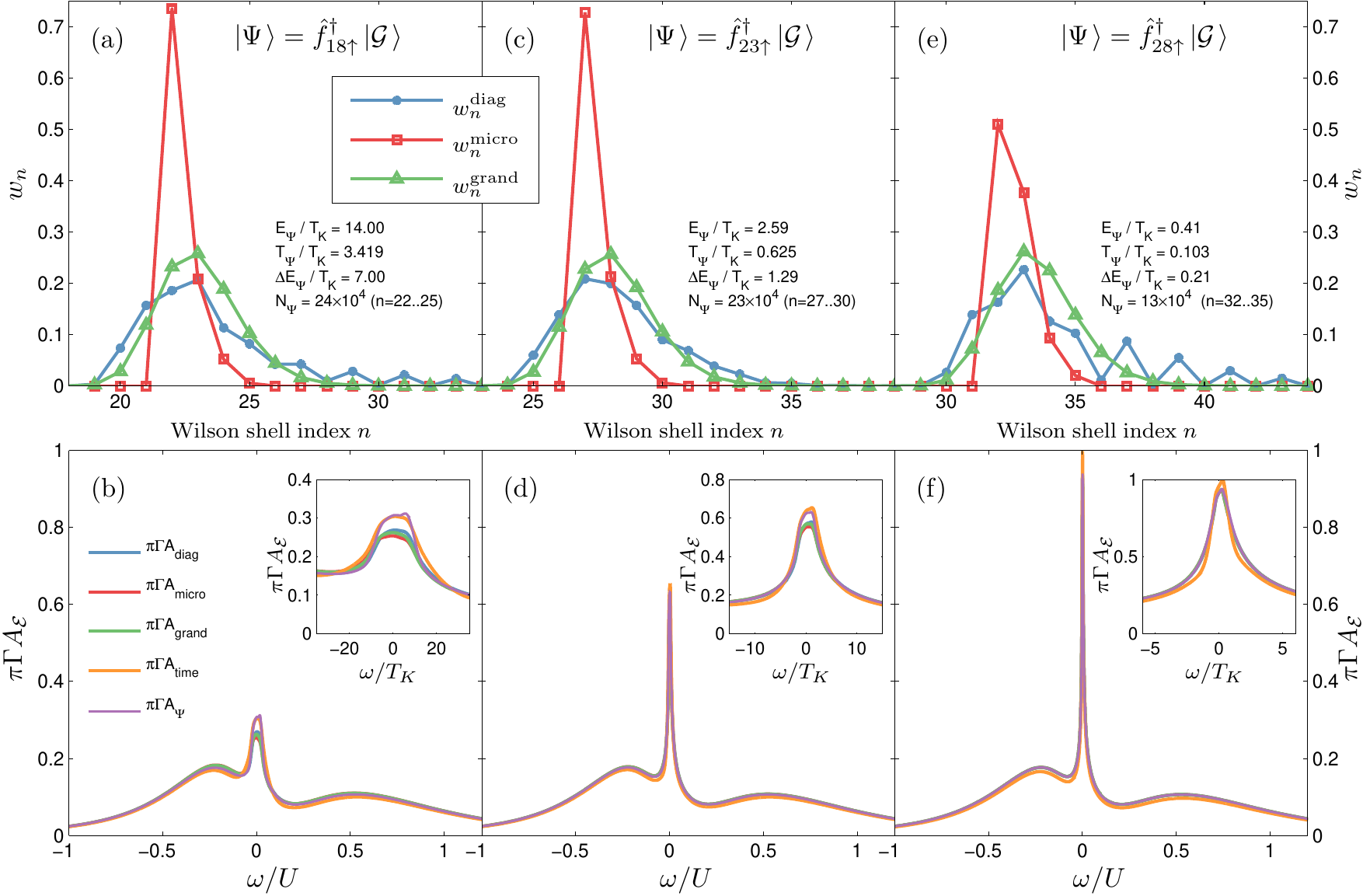}
  \caption{ \label{Fig:A_for_fdagPsi} (color online)
  The density matrix weight distributions for the diagonal,
  microcanonical and grand canonical ensembles as a function of the
  Wilson shell index $n$ (a,c,e), and the normalized spectral
  functions (b,d,f) calculated for a few
  excited states obtained after acting with the spin-up creation
  operator of shell $n$, $\hat{f}_{n\uparrow}^\dagger$, on the ground
  state: (a,b) $\ket{\Psi}=\hat{f}_{18\uparrow}^\dagger
  \ket{\mathcal{G}}$, (c,d) $\ket{\Psi}=\hat{f}_{23\uparrow}^\dagger
  \ket{\mathcal{G}}$, and (e,f)
  $\ket{\Psi} = \hat{f}_{28\uparrow}^\dagger \ket{\mathcal{G}}$.
  The number of states contributing to the
  microcanonical ensemble is denoted by $N_\Psi$, where
  the notation $n=(n_1..n_2)$
  indicates the contributing shells.
  The insets present the zoom of the spectral functions
  around the Kondo peak.
}
\end{figure*}

To begin with and also for later reference, let us start with
analyzing the behavior of the normalized spectral function
$\pi\Gamma A_{\mathcal{G}}$ for the ground state of the system, \ie
$\ket{\Psi} = \ket{\mathcal{G}}$, where through Friedel-sum-rule,
one expects $\pi\Gamma A(\omega=0) \simeq 0.99$ (see
Fig.~\ref{Fig:A_for_GS}).
The spectral function displays two broad maxima at resonant energies
$\w = \e_d$ and $\w = \e_d+U$, and a narrow Kondo resonance
\cite{hewson_book} of width $T_K$ pinned at the Fermi level $\w=0$.

The spectral functions shown in Fig.~\ref{Fig:A_for_GS} are
calculated by using the diagonal ensemble, $A_{\rm diag}$, the
microcanonical ensemble, $A_{\rm micro}$, and grand canonical
ensemble, $A_{\rm grand}$.
Because $\ket{\Psi} = \ket{\mathcal{G}}$ is an eigenstate of the
Hamiltonian, the diagonal, microcanonical and grand canonical (at
temperature $T<\Lambda^{-N/2}$) density matrices have only a single
nonzero entry, \ie from the state, $\ket{se}^D_n =\ket{\mathcal{G}
}$. Consequently, by construction, one finds $A_{\rm diag} = A_{\rm
micro} = A_{\rm grand} = A_{\Psi}$. In principle, also the
time-averaged spectral function $A_{\rm time}$ should be exactly
equal to $A_{\Psi}$. However, as seen in the inset in
\fig{Fig:A_for_GS}, here some small differences arise from the fact
that to calculate $A_{\rm time}$, cf. \eq{Eq:At2}, one needs to
insert one additional completeness relation (\ref{Eq:closrel}) as
compared to the calculation of $A$ from a density matrix. Due to the
NRG approximation in \eq{eq:NRG:approx}, indeed, this can lead to
slightly different numerical results.

\subsection{Single-particle excitations in the bath
\label{sec:x:1particle}}

While the eigenstate thermalization hypothesis is trivially
satisfied for the ground state of the system
(see \fig{Fig:A_for_GS} above),
the validity of ETH is not at all clear for other
eigenstates of the system. 
In the following, more generally, we analyze thermalization
by starting from pure states which are not necessarily also
eigenstates of the full Hamiltonian.
Here, in particular, we consider excited states of the bath 
obtained by adding or removing a single particle at energy shell $k$
within the bath on top of the Kondo ground state. To be specific,
we consider $\ket{\Psi}=\hat{f}_{k\uparrow}^\dagger
\ket{\mathcal{G}}$ with $k=18,\,23,\,28$
for the same model parameters as in \fig{Fig:A_for_GS}.
The energy of such an excitation is approximately
proportional to $ \omega_k \simeq D\Lambda^{-k/2}$. For the
parameters chosen here, shell $k =n_K \simeq 23$ corresponds to the
energy scale of the Kondo temperature $T_K$, whereas $k=18$ ($k=28$)
corresponds to a larger (smaller) energy scale, respectively. The
spectral functions calculated using different ensembles are
displayed in \fig{Fig:A_for_fdagPsi}. The top panels present
the dependence of weights of the corresponding ensembles on the
Wilson shell index $n$, while the bottom panels show the
corresponding spectral functions.

It is instructive to compare the shapes of the spectral
functions of these excited states to that of the ground state
spectral function in \fig{Fig:A_for_GS}.
Since the energies of these excited states are still
significantly smaller than the energies corresponding to the Hubbard
resonances, $E_\Psi \ll |\e_d|,\e_d+U$, the high-energy features of
the spectral functions remain essentially
unaltered. What is most affected for
given parameter set, is the height of the resonance at the Fermi
level, which is a measure of the respective strength of electronic
correlations leading to the Kondo effect. As expected, the overall
trend seen in the lower panels of \fig{Fig:A_for_fdagPsi}
is that for $k < n_K$, the Kondo resonance is significantly
suppressed [see Fig.~\ref{Fig:A_for_fdagPsi}(b)], for $k = n_K$, the
normalized spectral functions reach approximately one half [see
Fig.~\ref{Fig:A_for_fdagPsi}(d)], while for $k > n_K$, the Kondo
resonance almost reaches its maximum possible height [see
Fig.~\ref{Fig:A_for_fdagPsi}(f)]. In other words, hitting the ground
state with an operator acting on energy shell $k$ destroys
low-energy correlations at energies $\omega_{n} \ll \omega_k$, \ie $n>k$.

\begin{figure}[t]
\includegraphics[width=1\columnwidth]{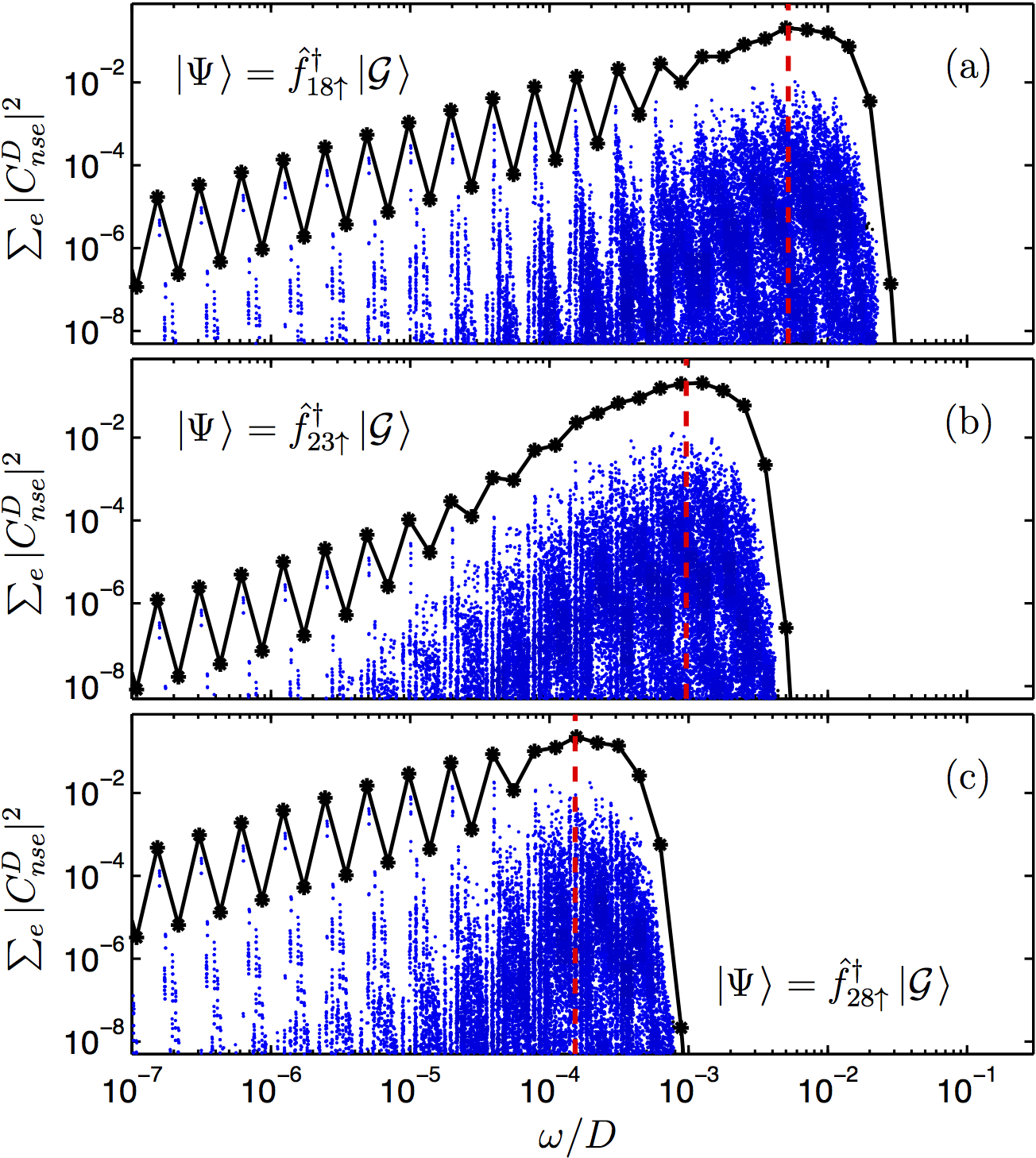}
  \caption{\label{Fig:Cns_logE} (color online)
  The matrix elements of the diagonal density matrix, 
  $\sum_e |C_{nse}^D|^2$,
  calculated for different excited states, $\ket{\Psi} =
  \hat{f}_{n\uparrow}^\dag \ket{\mathcal{G}}$, as indicated in the
  figure panels. The matrix elements are plotted as a function of energy
  $\omega\equiv E_{ns}^D$ measured relative to the ground state energy
  $E_\mathcal{G}$. The solid lines display the integrated weights
  $w_n^{\rm diag}$ of the diagonal ensemble plotted as a function of
  energy of a given iteration $n$, $\omega = \alpha
  \Lambda^{-n/2}$, with $\alpha$ a numerical constant
  taking into account the energy spread of each iteration.
  The vertical dashed lines indicate the average energy of state
  $\ket{\Psi}$, $E_\Psi$.
}
\end{figure}

Let us now focus more carefully on the particular behavior of $A$
for different ensembles and the implications of our numerical
results for the ETH. The first row of Fig.~\ref{Fig:A_for_fdagPsi}
shows the corresponding weights $w_n^{\mathcal{E}}$ [see
\eq{Eq:rho}] of the diagonal, microcanonical and grand canonical
ensembles. For a given state $\ket{\Psi}=\hat{f}_{k\uparrow}^\dagger
\ket{\mathcal{G}}$, throughout, the normalized weights become
nonzero at energy shells $n\gtrsim k$, exhibit a pronounced maximum
around $n \sim k+4$, which is followed by a rapidly decaying tail
towards larger $n$, \ie smaller energy scales. The shift of the
maximum in $w_{n}$ relative to $k$
depends on $\Lambda$, and
results from the fact that although the energy of state $E_{\Psi}$
is comparable to the energy scale of given iteration $n$, the
representative states relevant for the microcanonical ensemble,
satisfying $|E_{ns}^D - E_\Psi |\leq \delta E_\Psi$, belong to
iterations $n>k$. This also underpins the choice for the effective
temperature $T_\Psi \simeq E_\Psi/4$ for the grand canconical
ensemble which ensures that $n_{\rm max}^{\rm grand} \simeq n_{\rm
max}^{\rm diag}$.

By construction, the weight distributions $w^{\rm micro}_{n}$ are
significantly narrower, yet also with their maximum at $n_{\rm
max}^{\rm micro} \sim k+4$. The number of Wilson shells relevant for
the microcanonical ensemble are comparable for all states considered
in Fig.~\ref{Fig:A_for_fdagPsi}(a,c,e). In contrast, the
weight distributions $w_n^{\rm diag}$ and $w_n^{\rm grand}$
for the diagonal and grand canonical ensemble, respectively, are
nonzero over a wider range of shells and spread to the end of chain.
Nevertheless, the maximum weight occurs at a comparable energy
shell for all these cases,
and therefore $n_{\rm max}^{\rm diag} \simeq n_{\rm max}^{\rm
micro} \simeq n_{\rm max}^{\rm grand}$. Because of that, all the
three ensembles give comparable results, which become basically
identical once $k > n_K$ [see Fig.~\ref{Fig:A_for_fdagPsi}(f)].
Therefore with $A_{\rm diag} \simeq A_{\rm micro}$, this suggests
that for the initial states with single-particle excitations in the
bath, $A$ behaves thermally and, in the long time limit, can be
described by a proper statistical-mechanical ensemble.

The detailed energy dependence of the matrix elements of the
diagonal density matrix $\sum_e |C^D_{nse}|^2$ \vs $E_{ns}^D$ is presented
in Fig.~\ref{Fig:Cns_logE} for the same three excited states
as analyzed in Fig.~\ref{Fig:A_for_fdagPsi}. 
The overall behavior of the matrix elements
$|C^D_{nse}|^2$ is reflected in the integrated weights
resulting in the diagonal density ensemble $w_n^{\rm diag}$
(solid lines with bullets).
Its distribution is clearly peaked around the average energy  $E_\Psi$
(red-dashed vertical lines in \fig{Fig:Cns_logE}),
with a \emph{linear} decay towards lower
energies. The physical relevance of the latter will be
analyzed in detail in \Sec{sec:x:quench} and \fig{Fig:weights} below.

\begin{figure}[t]
\includegraphics[width=0.9\columnwidth]{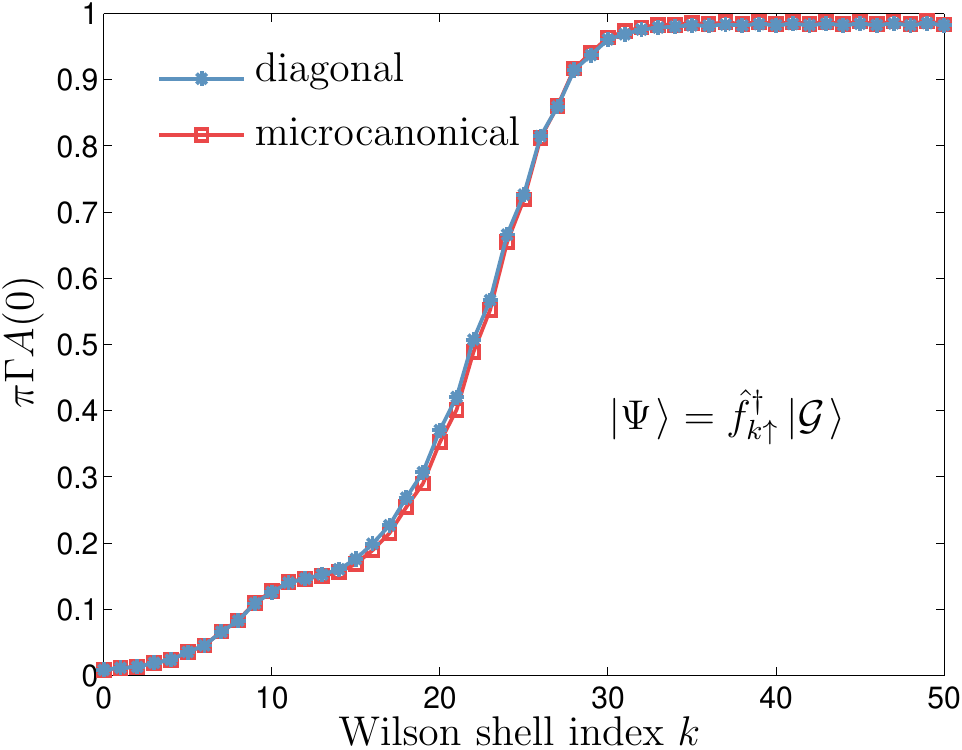}
  \caption{\label{Fig:A0_fdagPsi} (color online)
  The normalized spectral function at energy $\omega=0$, $\pi\Gamma
  A(0)$, calculated using the diagonal and microcanonical ensembles
  for the initial state $\ket{\Psi} = \hat{f}_{k\uparrow}^\dag
  \ket{\mathcal{G}}$ as a function of the Wilson shell index $k$.
  The spectral data was broadened with $\omega_0 = E_\Psi$. The
  predictions based on microcanonical ensemble agree reasonably well
  with those based on the diagonal ensemble, which indicates that the
  system behaves thermally.
}
\end{figure}

The long time behavior of the system is
approximately described by the microcanonical ensemble.
This is demonstrated in \fig{Fig:A0_fdagPsi}
which shows the dependence of the
height of the Kondo resonance, $\pi\Gamma A(0)$, on the Wilson shell
index $k$ for the initial state $\ket{\Psi} =
\hat{f}_{k\uparrow}^\dag\ket{\mathcal{G}}$. This static observable
can be directly related to the linear conductance at given
temperature, $G/G_0 = \pi\Gamma A(0)$, with $G_0 = 2e^2 / h$. One
can see that with increasing $k$, $\pi\Gamma A(0)$ shows a small
resonance for $k\approx 10$, which corresponds to the energy scale
of the hybridization $\Gamma$. 
This linear conductance increases further all the way
down to the Kondo energy scale $n_K \simeq23$.
For $k > n_K$, finally, $\pi\Gamma A(0)$ saturates and reaches unity.
The fact that $A_{\rm diag} \simeq A_{\rm micro}$ holds, \cite{Rigol_Nature08}
supports the eigenstate thermalization hypothesis,~\cite{Deutsch,Srednicki} which states
that the eigenstate expectation values do not fluctuate between
eigenstates that are close in energy.

\begin{figure}[t]
\includegraphics[width=0.99\columnwidth]{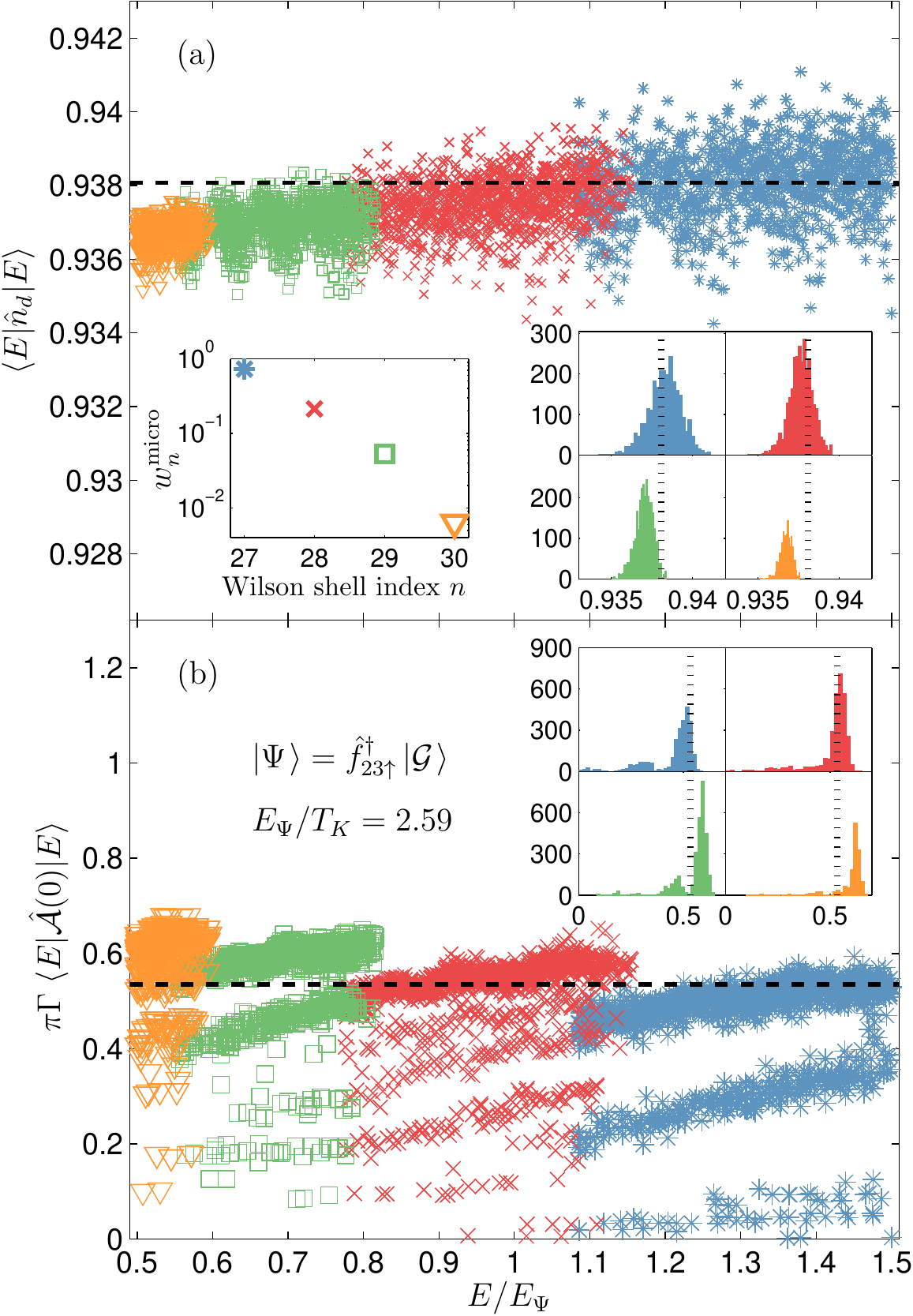}
  \caption{\label{Fig:ETH_fdagPsi} (color online)
  Energy-resolved expectation values relevant for the microcanonical
  ensemble of (a) the dot occupation, $n_{d} =\langle E |
  \hat{n}_{d} | E \rangle$, and (b) the normalized spectral
  function operator $ \pi\Gamma \langle E | \hat{\mathcal{A}}(0) | E
  \rangle$ taken at $\omega=0$. Here the microcanonical ensemble was
  based on the initial state $\ket{\Psi} = \hat{f}_{23\uparrow}^\dag
  \ket{\mathcal{G}}$, and the energy expectation values are computed
  for individual energy eigenstates $\ket{E}\equiv \ket{se}_n^D$. For
  (b), the spectral data was broadened using $\omega_0 = E_\Psi$. The
  dashed horizontal lines display ${\rm Tr}[\hat{\rho}_{\rm micro}
  \hat{n}_{d} ]$ and $A_{\rm micro}$ for panel (a) and (b),
  respectively. The left inset in (a) presents the distribution of
  weights of the microcanonical density matrix $w_n^{\rm micro}$ on
  the Wilson chain, where the different symbols differentiate between
  specific energy shells. The same symbols are also used when plotting
  the energy expectation values in the main panels. The right insets
  in (a) and (b) show the corresponding energy-shell resolved
  histograms of the data from the main panel.
}
\end{figure}

To check whether ETH really holds for our strongly correlated
electron system, instead of the above single-particle
excited states, we also analyze the behavior of the actual
many-body eigenstates. For this purpose,
\fig{Fig:ETH_fdagPsi} presents the energy
expectation values of two specific physical quantities for
individual NRG eigenstates $\ket{E}\equiv
\ket{se}_n^D$ relevant for the microcanonical ensemble, \ie satisfying
$|E-E_\Psi| \leq \delta E_\Psi$
for some given reference state $\ket{\Psi}$:
\fig{Fig:ETH_fdagPsi}(a) shows the plain dot occupation
$n_{d} \equiv \langle E | \hat{n}_{d} | E \rangle$ with
$\hat{n}_{d} = \sum_\sigma \hat{d}^\dag_\sigma\hat{d}_\sigma$,
where in the thermal case $n_{d}= \int d\omega A(\omega)
f(\omega)$ with $f(\omega)$ the Fermi distribution function, and
\fig{Fig:ETH_fdagPsi}(b) depicts
the normalized spectral function operator $\hat{\mathcal{A}}$
taken at energy $\omega=0$, \ie $\pi\Gamma \, \langle E |
\hat{\mathcal{A}}(0) | E \rangle$.

The expectation values are calculated for the 
reference state
$\ket{\Psi} = \hat{f}_{23\uparrow}^\dag\ket{\mathcal{G}}$
where the single-particle excitation occurs at the energy shell
$k=n_K\simeq 23$ corresponding to $T_K$. They are compared with the
microcanonical expectation values, ${\rm Tr}[\hat{\rho}_{\rm micro}
\hat{n}_{d} ]$ and $A_{\rm micro}$, respectively (see the
dashed horizontal lines in \fig{Fig:ETH_fdagPsi}). The different
colors (symbols) indicate those Wilson shells at which the
weights of the microcanonical density matrix are finite [see left
inset in \fig{Fig:ETH_fdagPsi}(a)]. 
The largest contribution comes from iteration $n=27$,
where $w_n^{\rm micro}$ is maximum.

Let us first discuss the behavior of the plain expectation
value of the dot occupation $n_{d}$ in
\fig{Fig:ETH_fdagPsi}(a). While it
grows weakly with increasing $E$, the histogram
of the data in the right inset of \fig{Fig:ETH_fdagPsi}(a) clearly
shows that the energy expectation values are centered at the
microcanonical expectation value (vertical dashed lines).
The spread of the data is extremely small due to the simplicity of the
measured operator which, by itself, is insensitive to Kondo
correlations. Overall, however, this clearly indicates that for
given operator the knowledge of a single energy-eigenstate within a
narrow energy window is sufficient to find the expectation value
$\expect{\hat{n}_{d}}$ in the long-time limit.

In contrast, the energy expectation value of the normalized spectral
operator at the Fermi level, $\pi\Gamma \hat{\mathcal{A}}(0)$, is
significantly more sensitive to the specific energy eigenstate, as
shown in \fig{Fig:ETH_fdagPsi}(b). The reason for this is that in
given case, by construction, $\pi\Gamma\hat{\mathcal{A}}(0)$
derives from a dynamical correlator that
is strongly sensitive to Kondo correlations. In \fig{Fig:ETH_fdagPsi}(b), due to the discrete
nature of the Wilson chain at intermediate iteration $n$, the data
appears in bunches. Similar to \fig{Fig:ETH_fdagPsi}(a), each of
these bunches shows a slight upward trend with increasing energy
$E$, while the overall trend from shell to shell is slightly
downward. Nevertheless, although the fluctuations of the data are larger
than in \fig{Fig:ETH_fdagPsi}(a), the main contribution is
centered at the microcanonical expectation value, $A_{\rm micro}$,
as seen in the insets to \fig{Fig:ETH_fdagPsi}(b). This again
suggests that the ETH can be invoked as the mechanism responsible
for thermalization in the considered system. The knowledge of an
expectation value for a representative energy eigenstate, \ie an
energy eigenstate $\ket{E}$ around $E_\Psi$ satisfying $|E-E_\Psi|
\leq \delta E_\Psi$, can be thus used to predict the behavior of the
system in the long-time limit.

\subsection{Density excitations in the bath
\label{sec:x:density}}

\begin{figure}[t]
\includegraphics[width=0.9\columnwidth,clip]{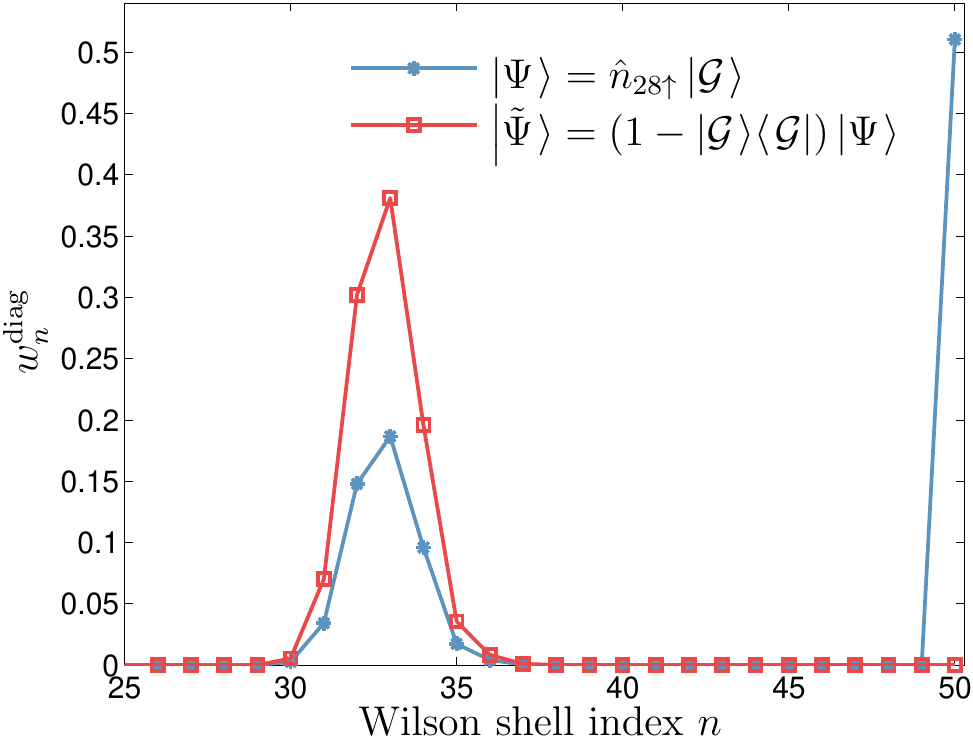}
  \caption{ \label{Fig:weights_densexc}
  (color online) The weights of the diagonal ensemble $w_n^{\rm diag}$
  as a function of the Wilson shell index $n$ calculated for states
  $\ket{\Psi} = \hat{n}_{28\uparrow} \ket{\mathcal{G}}$ and
  $\ket{\tilde \Psi} = (1-\ket{\mathcal{G}}\bra{\mathcal{G}})
  \ket{\Psi}$.
}
\end{figure}

We now study the process of thermalization and time evolution for
initial states obtained after acting with a single-particle {\it
density} operator onto the ground state of the system, $\ket{\Psi}=
\hat{n}_{k\uparrow} \ket{\mathcal{G}}$ with $\hat{n}_{k\uparrow}
\equiv \hat{f}_{k\uparrow}^\dagger \hat{f}_{k\uparrow}
^{\phantom{\dagger}}$. This corresponds to a projection
of the ground state $\ket{\mathcal{G}}$ onto its component
where site \mbox{$(k\!\uparrow)$} is fully occupied.
The weights of the diagonal density matrix
$w_n^{\rm diag}$ for $k=28$ are shown in \fig{Fig:weights_densexc}.
The distribution of $w_n^{\rm diag}$ has a maximum for iterations
slightly larger than the Wilson shell on which the density operator
acts, yet also has a delta-like contribution at the very end of the
finite Wilson chain of length $N$ considered. In fact, approximately
half of the weight is transferred to the last shell $n=N$. This can
be understood by realizing that for given case $w_n^{\rm diag} =
\sum_{se}|{}^D_n\!\bra{se} \hat{n}_{k\uparrow}  \ket{\mathcal{G}}
|^2$ has a singular static contribution for $\ket{se}_n =
\ket{\mathcal{G}}$, namely $|\bra{\mathcal{G}} \hat{n}_{k\uparrow}
\ket{\mathcal{G}}|^2 \simeq 1/4$ due to half-filling which, after
including the normalization of $\ket{\Psi}$ above, becomes
$\sim1/2$. This simple static contribution needs to be dealt with
separately. The remainder gives rise to a broad distribution around
the excitation energy.

For the discussion of statistical ensembles, we focus on the latter
term only. For this, we project out the ground state and use the
state $\ket{\tilde\Psi} = (1- \ket{\mathcal{G}}\bra{\mathcal{G}} )
\ket{\Psi} $ to study the process of thermalization [in practice,
this state $\ket{\tilde\Psi}$ can be constructed exactly by
orthonormalizing both states, $\ket{\Psi}$ as well as $\langle
\mathcal{G} | \Psi \rangle \ket{\mathcal{G}}$ towards the Wilson
shell $k$ at which the excitation was created, and then perform the
orthonormalization of $\ket{\Psi}$ with respect to representation of
$\ket{{\mathcal{G}}}$ there]. The resulting weight distribution for
state $\ket{\tilde\Psi}$ is shown in \fig{Fig:weights_densexc}. This
now again resembles the distribution of a thermal ensemble. We have
studied the expectation values of the spectral operator for state $\ket{\tilde\Psi} = (1-
\ket{\mathcal{G}}\bra{\mathcal{G}} ) \hat{n}_{k\uparrow}
\ket{\mathcal{G}} $ as a function of the Wilson shell index $k$ (not
shown), and found that again for all $k$ the microcanonical ensemble
describes very well the behavior of the system in the long-time
limit.

\subsection{Relevance of microscopic quantum quenches
and Anderson orthogonality
\label{sec:x:quench}}

The ensembles in \Sec{sec:x:1particle}
(\fig{Fig:A_for_fdagPsi}) were constructed from a
single particle excitation onto the ground state of the system.
Foremost, this corresponds to a microscopic quantum quench at zero temperature.
The spectral functions $A_{\rm time}$ and $A_\Psi$
analyzed in \fig{Fig:A_for_fdagPsi}, are based on the states
$\ket{\Psi}$ with or without time-evolution with respect to the full
Hamiltonian, respectively. Nevertheless, after real-time
evolution to infinite times (see also diagonal ensemble) the Kondo
resonance at $\w=0$ is not fully restored 
with only a slight difference between $A_\Psi$ and $A_{\rm time}$
[Figs.~\ref{Fig:A_for_fdagPsi}(b) and
\ref{Fig:A_for_fdagPsi}(d)].
This is essentially due to the fact that the Hilbert space, although
exponentially large, is based on a coarse-grained bath through the
logarithmic discretization parameter $\Lambda$ of NRG. Consequently,
the energy of a microscopic excitation cannot be fully
dissipated to infinity, and one can
only resolve the physics at energy approximately equal or larger
than the corresponding excitation energy. This is in agreement with
the simple fact \cite{rosch12} that Wilson chains are
not true thermal reservoirs and as such cannot fully transfer local
microscopic energy to infinity. In particular, due to the lack of
real-space association within the Wilson chain, energy
cannot be dissipated to the end of the chain
since for a given Wilson shell $n$, all shells $n'>n$ represent
and thus can absorb only exponentially small energy
$\propto \Lambda^{-n/2}$.
Therefore due to the exponential decrease of hopping
matrix elements along the Wilson chain, the energy of
an excitation that is created at a given site cannot
travel away very far in either direction along the
Wilson chain, since there is an energy mismatch in
both directions [(see also \fig{Fig:occup} below).

A microscopic local quantum quench within a Fermi sea
can also be analyzed through the viewpoint of Anderson orthogonality
(AO).\cite{anderson67,wb11aoc,munder11} The decay of the overlap of
ground state wave functions for initial (I) and final (F) state,
$\ket{\Psi_I}$ and $\ket{\Psi_F}$, respectively,
is given by\cite{anderson67}
\begin{subequations}\label{eq:AO-basics}
\begin{eqnarray}
 && | \langle \Psi_I | \Psi_F\rangle |^2
\propto
  \Bigl(\frac{1}{L}\Bigr)^{-\Delta n_\mathrm{loc}^2}
\label{eq:AO-basics:1}
\end{eqnarray}
where \cite{wb11aoc}
\begin{eqnarray}
  \frac{1}{L} \propto
   \omega_{(n)} \propto \Lambda^{-\tfrac{n}{2}} 
\label{eq:AO-basics:2}
\text{,} 
\end{eqnarray}
\end{subequations}
with $L$ the system size, and
$\frac{1}{L} \propto \omega_{(n)}$ the finite
size level-spacing. Hence in the NRG context, for a 
Wilson chain of linear length $n$ translates into
an {\it exponentially} large (effective) system sizes.
Therefore the AO power law decay in \eq{eq:AO-basics:1}
translates into an exponential decay along the
Wilson chain. \cite{wb11aoc}
When plotted vs. the energy scale $\omega_n$, the
fitted exponent reflects the charge $\Delta n_\mathrm{loc}$
that is displaced to or from infinity. For example,
the exponent seen in \fig{Fig:Cns_logE} for small
energies $\omega$ suggests
$|\Delta n_\mathrm{loc}|=1$, which is in perfect
agreement with the fact that the extra particle
$\hat{f}_{n\sigma}^\dag$ that has been inserted in
$|\Psi_I\rangle =
  \hat{f}_{n\uparrow}^\dag \ket{\mathcal{G}}$
will be dissipated to infinity.

The decay of the wave function overlap in \eq{eq:AO-basics:1},
is not necessarily constrained to the overlap of explicit
ground state wave functions. In particular, it
describes an arbitrary
local quantum quench,\cite{munder11} say at time $t=0$,
on top of the ground state of given (final) Hamiltonian
[$\ket{\Psi(t=0)}=\ket{\Psi_I}$ can be considered the ground state
of a fictitious initial Hamiltonian].
As time proceeds, a larger and larger volume $L$
will be affected through the quantum quench, which
justifies the association that $L \sim v_F t \propto 1/\omega$,
with $v_F$ the Fermi velocity.

Furthermore, the overlap in \eq{eq:AO-basics:1}
can be computed with respect to an ensemble average
at given energy scale $\omega_n \propto 1/L$,\cite{wb11aoc}
and thus the l.h.s. of \eq{eq:AO-basics:1} directly relates
to the diagonal weight distributions $w_n^\mathrm{diag}$.
\begin{eqnarray}
   w_{n;F}^\mathrm{diag} \propto
  \rho_{n;I}^\mathrm{diag} \cdot \Bigl(\frac{1}{L}\Bigr)^{-\Delta n_\mathrm{loc}^2}
\label{eq:AO:rho}\text{,}
\end{eqnarray}
where $\rho_{n;I}^\mathrm{diag} \equiv
\sum_{n'\geq n}  w_{n';I}^\mathrm{diag}$ describes the cumulative
weight of the diagonal density matrix of the initial system
for shells $n'\geq n$.
Therefore adding or removing a particle
strongly affects the
weight distributions $w_n^\mathrm{diag}$ of the final state.
This is the underlying mechanism that essentially allows
the very different earlier interpretation above of a (close
to) thermalized state, despite having applied a microscopic single
particle excitation. There
the resulting state mimics a macroscopic
ensemble while caveats apply (see \Sec{sec:thermolim} below). All of
this, of course, is very specific and hence tightly connected
to the underlying logarithmic discretization and hence to the NRG perspective.

\begin{figure}[t]
\includegraphics[width=1\columnwidth]{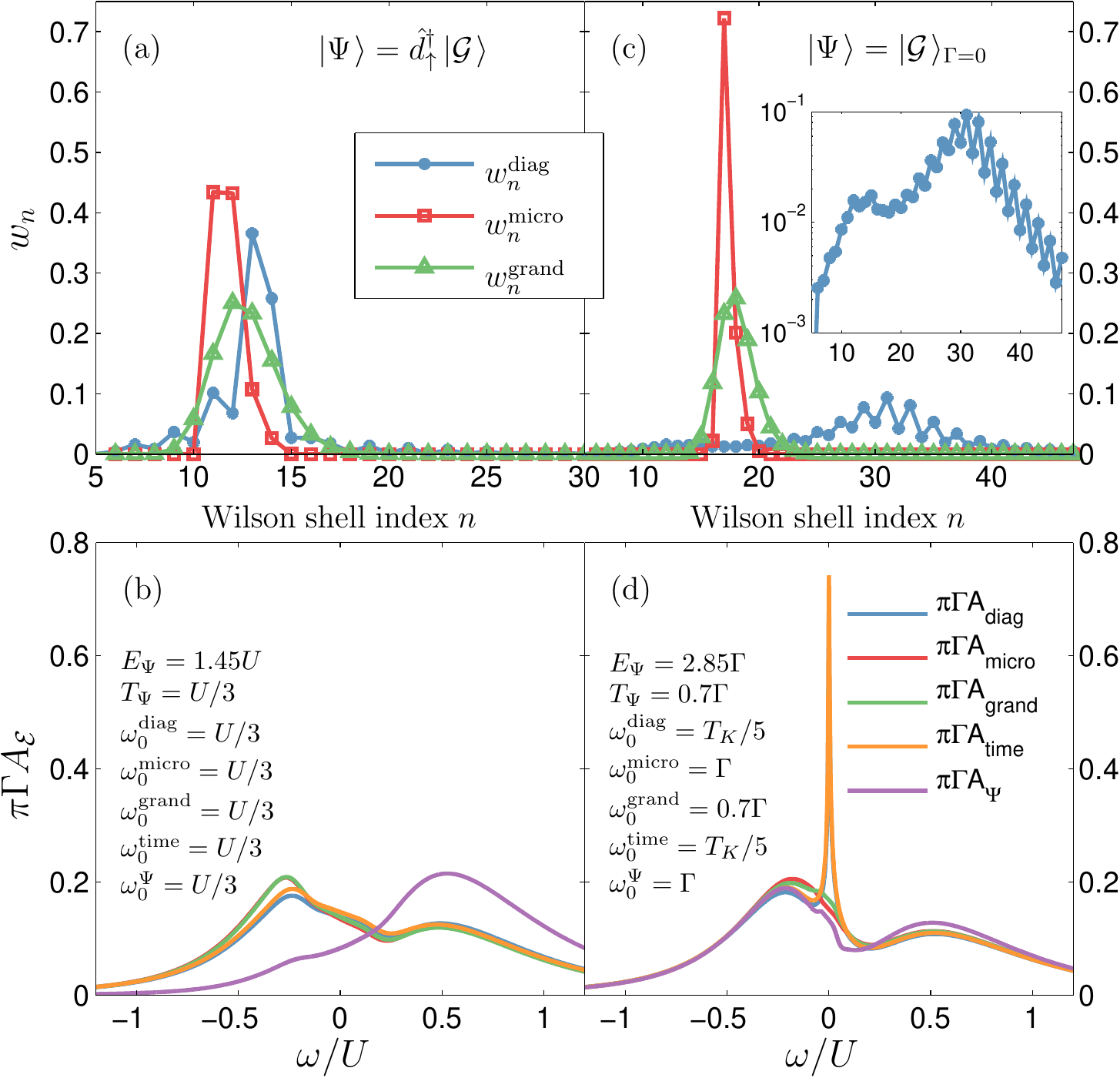}
  \caption{ \label{Fig:Adot} (color online)
  Weight distribution of various ensembles (upper panels) and the
  corresponding normalized spectral function $\pi\Gamma A_\mathcal{E}$ (lower
  panels) for the state $\ket{\Psi} = \hat{d}^\dag_{\uparrow} \ket{\mathcal{G}}$
  (left panels) and $\ket{\Psi} = \ket{\mathcal{G}}_{\Gamma=0}$ (right panels).
  The grand canonical spectral
  function was calculated at an effective temperature $T_\Psi$
  specified in the figure, together with $\omega_0^{\rm grand} =
  T_\Psi$. For the other spectral functions, $A_{\rm diag}$, $A_{\rm
  micro}$, $A_{\rm time}$, and $A_\Psi$,
  we optimized the broadening parameters such
  that all unphysical features due to discretization were smeared out,
  with the respective values for
  $\omega_0^{\rm diag}$, $\omega_0^{\rm
  micro}$, $\omega_0^{\rm time}$, and $\omega_0^{\Psi}$ 
  specified in the panels. The inset to panel (c) shows
  the weights of the diagonal ensemble $w_n^{\rm diag}$ plotted on logarithmic scale
  as a function of Wilson shell index.
}
\end{figure}

While in the above sense, \fig{Fig:A_for_fdagPsi} does allow an
approximate effective interpretation of single-particle excitations
within the Wilson bath in terms of effective thermodynamic ensembles
(a more careful analysis will be given in \Sec{sec:thermolim}
below), clearly, this interpretation is not always possible. 
Motivated by above AO based arguments,
in the following we contrast two local quantum quenches at the impurity:
(i) using $\ket{\Psi} = \hat{d}^\dag_{\uparrow}\ket{\mathcal{G}}$, and
(ii) using the ground state of the decoupled bath, $\ket{\Psi} =
\ket{\mathcal{G}}_{\Gamma=0}$, which is equivalent to a quantum
quench by turning on $\Gamma$. The former inserts a particle,
while the latter turns on the hybridization of the impurity
which mostly results in only local rearrangement of charge.

Figure~\ref{Fig:Adot} presents the weight distributions of the
corresponding ensembles and the spectral functions.
The analysis of the state $\ket{\Psi_b} =
\hat{d}^\dag_{\uparrow}\ket{\mathcal{G}}$ [\fig{Fig:Adot}(a,b)]
is similar in spirit to the analysis in \fig{Fig:A_for_fdagPsi},
with the minor difference that
the single-particle excitation does not occur within the
logarithmically discretized bath through the application of some
operator $\hat{f}^\dag_{k\uparrow}$ which is delocalized in real
space, but rather through the application of
$\hat{d}^\dag_{\uparrow}$ which acts locally
at the impurity itself.
Therefore the initial spectral function $A_\Psi$
shows significant shift of spectral weight from the
lower towards the upper Hubbard resonance at $\omega/U \simeq 0.6$
since by creating an extra particle in the dot enhances the double occupancy.
Over time, however, this again relaxes back to the lower
Hubbard resonance [see \fig{Fig:Adot}(b)].
Furthermore, by having adding a particle, its charge eventually
has to be dissipated to infinity. Hence 
similar to \fig{Fig:A_for_fdagPsi},
starting from the energy scale of the excitation, i.e.
quantum quench, strong AO effects
again cut off the weight distribution towards later Wilson shells.
Therefore all the weight distributions and the spectral functions
for the different statistical ensembles show similar behavior,
including $A_{\rm diag} \simeq A_{\rm micro}$.

Now consider the dynamics starting from the decoupled quantum dot
in the state $\ket{\Psi} = \ket{\mathcal{G}}_{\Gamma=0}$ 
[\figs{Fig:Adot}(c,d)]. Since there are no Kondo correlations
in the initial state $\ket{\Psi}$, clearly, $A_\Psi$ shows no Kondo resonance at all.
Using an effective temperature such that
$E_\Psi=\trace( \hat{\rho}_{\rm grand}\hat{H} ) \simeq 2.85\,\Gamma$,
this leads to $T_\Psi \simeq 0.7\,\Gamma \gg T_K$. Therefore the Kondo resonance is clearly also
absent in the grand canonical spectral function. For the same
reason, the results for the microcanonical spectral function are
also very similar. As seen in \fig{Fig:Adot}(c), both have a maximum
in their weight distribution at early iterations which correspond to
the energy scale of $\Gamma$.

\begin{figure}[t]
\includegraphics[width=0.9\columnwidth]{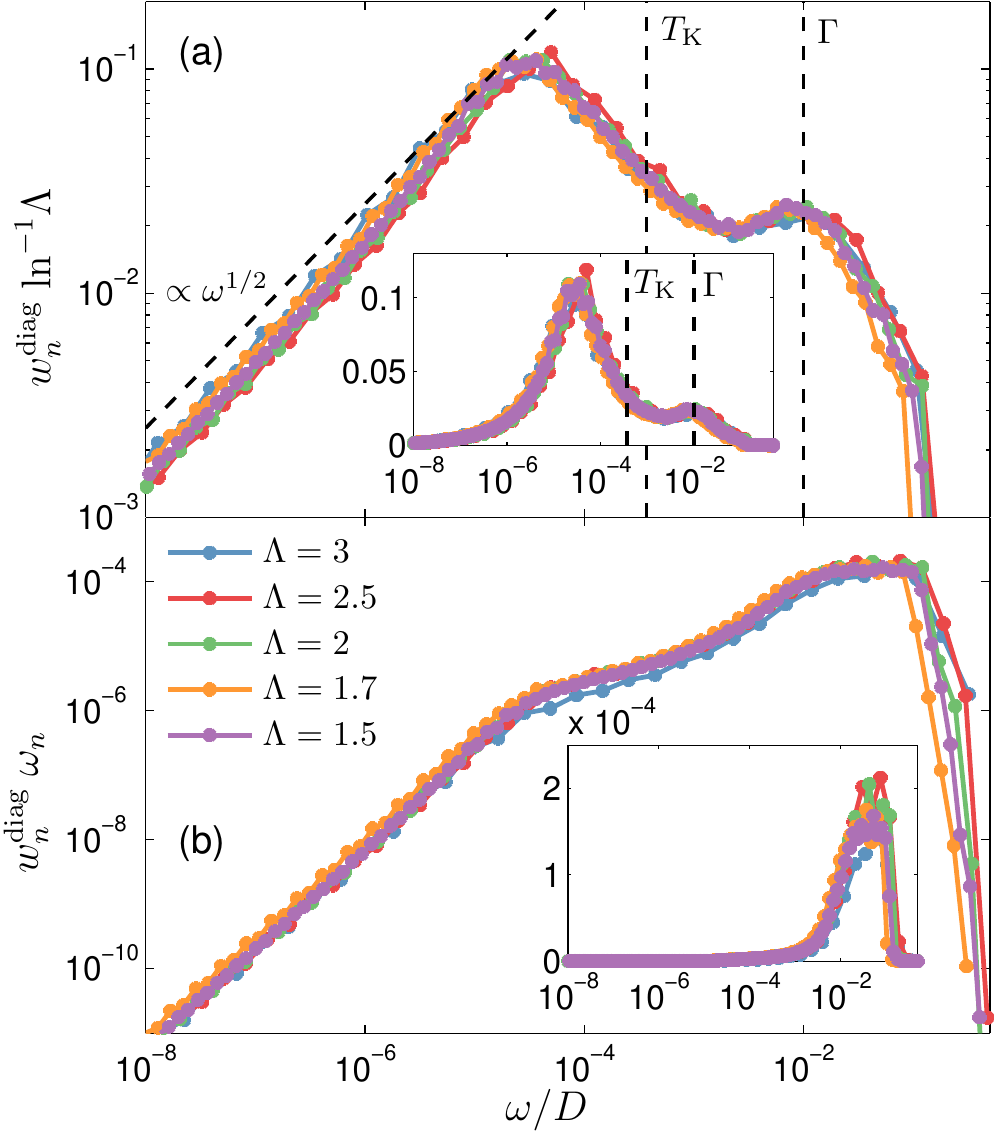}
  \caption{ \label{Fig:wndiagG0} (color online)
  (a) The weights of the diagonal density matrix $w_n^{\rm diag}$
  and (b) the total energy $E_\Psi \simeq w_n^{\rm diag} \; \omega_n$
  calculated for the state
  $\ket{\Psi} = \ket{\mathcal{G}}_{\Gamma=0}$  
  and plotted as a function of energy $\omega \propto \alpha_\Lambda \Lambda^{-n/2}$,
  with $\alpha_\Lambda$ constant of the order of unity chosen
  in such a way that the maximum of weight distribution
  occurs at the same energy for all values of $\Lambda$ presented in the figure.
  The insets show the corresponding data plotted on the $y$-axis linear scale.
  The vertical dashed lines in (a) and its inset
  mark the energy scales corresponding to 
  $\TK$ and $\Gamma$, respectively.
}
\end{figure}

In stark contrast, however, the weight distribution for the diagonal
ensemble in \fig{Fig:Adot}(c) has its maximum significantly
shifted to later energy shells, which clearly
lies beyond $n_K \simeq 23$.
As a direct consequence, the spectral data in panel (d)
clearly shows an emerging Kondo resonance. Overall, this results in a strong
mismatch $A_{\rm diag} \neq A_{\rm micro}$, which contradicts and
thus cannot be explained through ETH. 
Instead, here only the elementary description in terms of a
microscopic quantum quench in $\Gamma$ applies:
by dissipating the inserted microscopic energy to infinity,
from the point of view of the impurity
the system returns to its fully interacting ground state. 
In this view, however one
would expect that the Kondo resonance fully reemerges in the
thermodynamic limit. Nevertheless, in \fig{Fig:Adot}(d), a somewhat
suppressed Kondo peak is seen for $A_{\rm diag}$ as compared to
$T=0$ (see \fig{Fig:A_for_GS}).
Because $w_n^{\rm diag}$ is peaked around $T_K$,
the spectral data for smaller energies 
is limited due to the lack of spectral resolution for
$\omega \ll T_K$ for NRG specific technical reasons.
A more careful inspection of the
weight distribution $w_n^\mathrm{diag}$ [see inset to \fig{Fig:Adot}(c)] 
also reveals an enhanced weight around the energy scale of $\Gamma$ ($n\approx 13$).
This is required since for the non-interacting bath
under consideration,
even in the continuous i.e. non-discretized model,
particle or particle-hole excitations which are generated
close to the impurity at finite energy, can be emitted
to spatial infinity. There, however, they represent long-lived
particles due to the absence of interaction, and hence
no longer thermalize.
This is precisely also reflected within the NRG,
where the Wilson chain represents a closed yet exponentially
large system.
Being discrete, the energy $E_\Psi \sim \Gamma$ must be
contained around its respective energy shell.

The energy dependence of the diagonal weight distribution 
for different discretization parameter $\Lambda$
is explicitly shown in \fig{Fig:wndiagG0}(a).
The weights are plotted as a function of 
$\omega = \alpha_\Lambda \Lambda^{-n/2}$, with
$\alpha_\Lambda \sim 1$
being a constant of the order of unity
chosen such that the maximum in $w_n^\mathrm{diag}$ occurs
at the same energy for all values of $\Lambda$ considered in the figure.
Clearly, irrespective of $\Lambda$,
the structure of
the weight distribution 
 \emph{remains exactly the same} and features two peaks:
the first one occurs
for energies of the order of the coupling strength $\omega \approx \Gamma$,
while the second one is located around
$\omega \approx T_K$.
Since the total energy is given by
$E_\Psi \simeq w_n^{\rm diag} \; \omega_n$,
with $\omega_n \propto \Lambda^{-n/2}$, almost all energy
(more than 99\%) is carried by the weight for iterations
up to $n'\lesssim n_K$ ($\omega \lesssim T_K$), see
\fig{Fig:wndiagG0}(b).
The peak around $\omega \approx T_K$ and with it
the drop of the weight
distribution for $n>n_K$ ($\omega < T_K$), finally,
again can be understood through
Anderson orthogonality:
The initial state ($\Gamma=0$) has no Kondo
correlations at all, while the final state develops
a fully screened impurity spin, i.e. a Kondo singlet.
This is associated with a phase shift of $\pi/2$,
or equivalently, an effective shift in local charge
by half a charge for spin-up and spin-down, individually.
This is clearly also reflected in the AO exponent
derived for small energies $\omega$, as shown in
\fig{Fig:wndiagG0}(a).
In summary, in given case of 
$\ket{\Psi} = \ket{\mathcal{G}}_{\Gamma=0}$,
the effective temperature is therefore
\emph{not} set by $E_\Psi$. 
Rather, from the point of view of the impurity,
the system can be interpreted as having an effective
approximate temperature of $T^\ast_\Psi \lesssim T_K$
super-imposed with the single-particle excitations
that moved to spatial infinity. Therefore
$E_\Psi \gg T^\ast_\Psi$.
This is the reason why in this case
$A_{\rm diag} \neq A_{\rm micro}$.

\begin{figure}[t]
\includegraphics[width=0.9\columnwidth]{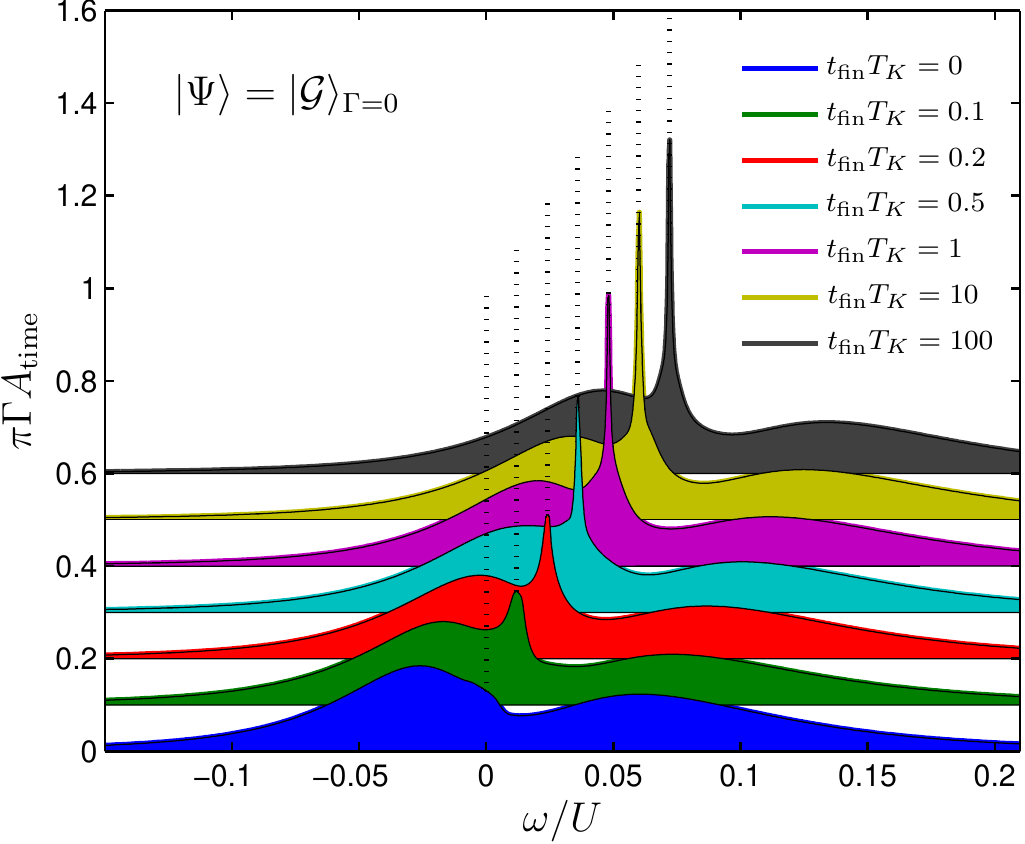}
  \caption{\label{Fig:A_for_different_time} (color online)
  The normalized spectral function for the time-evolved state $A_{\rm
  time}$ starting from
  $\ket{\Psi} = \ket{\mathcal{G}}_{\Gamma=0}$
 calculated at different final times $t_{\rm fin}$ as
  indicated in the figure. The spectral data was broadened with
  time-dependent broadening parameter, $\omega_0 = {\rm max}(1/t_{\rm
  fin}, T_K)$, with $\omega_0=\Gamma$ for $t_{\rm fin}=0$, and
  averaged over time interval $(t_{\rm fin}-\delta t,t_{\rm fin})$,
  with $\delta t = 0.2 t_{\rm fin}$. The spectral functions are
  shifted by $0.01$ ($0.1$) in the x-axis (y-axis) to increase
  visibility.
}
\end{figure}

The emergence of the Kondo resonance in the spectral function as a
function of time for $\ket{\Psi} = \ket{\mathcal{G}}_{\Gamma=0}$ at
$t=0$ is explicitly shown in Fig.~\ref{Fig:A_for_different_time}.
There $A_{\rm time}$ is calculated at different final times $t_{\rm
fin}$ using $\omega_0 =\max(1/t_{\rm fin},T_K)$. As expected, the
characteristic time scale, at which the Kondo correlations develop,
is of the order of $t_{\rm fin} \sim 1/T_K$. In the long-time limit,
the Kondo resonance saturates at the height $\pi\Gamma A_{\rm
time}(0) \sim 0.7$ which, indeed, is consistent with an effective
temperature $T_\Psi^\ast \lesssim T_K$. This long-time limit of
$A_{\rm time}$, by construction, is equivalent to the spectral
function $A_{\rm diag}$ [see \fig{Fig:Adot}(d)].

Summarizing \Sec{sec:x:1particle} to
\Sec{sec:x:quench}, microscopic quantum quenches first
and foremost need to be viewed from the perspective of Anderson
orthogonality. Their interpretation in terms of a thermalization to
a macroscopic statistical ensemble is peculiar to the logarithmic
discretization underlying NRG as
will be analyzed in more detail next.

\subsection{Thermodynamic limit
\label{sec:thermolim}}

Thermalization of a macroscopic system in the ground state to finite
temperature through application of a single-particle (written as
1-particle below) excitation appears like a contradiction.
Nevertheless, for standard discretization parameters (\ie $\Lambda
\gtrsim 2$), NRG comes very close to this description. On the other
hand, in the thermodynamic limit $\Lambda\to1^+$, this prescription
must fail and, eventually, macroscopically many 1-particle
excitations will be required for thermalization. While the location
of the maxima in the weight distribution $w_n$ along the Wilson
energy shells essentially will remain unaltered, the clear
difference will be in the tails of these weight distributions.
Therefore this section is devoted to the discussion of the behavior
of these tails for different discrete coarse-graining $\Lambda$, as
well as their behavior after subsequent multiple 1-particle
excitations towards exact thermalization. For the sake of energy
scale separation, however, in practice $\Lambda\gtrsim 1.4$ will be
used.

\begin{figure}[t]
\includegraphics[width=0.95\columnwidth,clip]{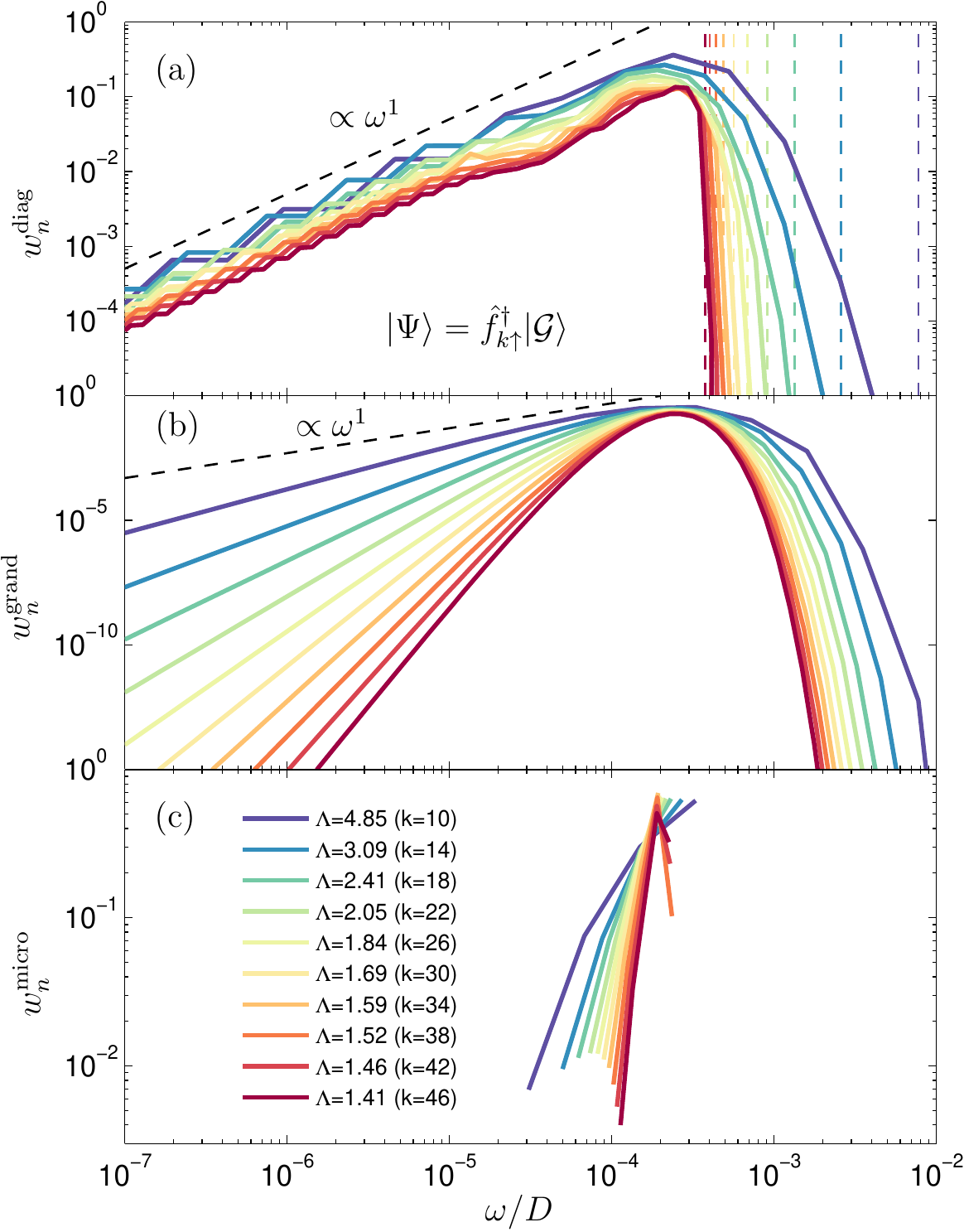}
  \caption{ \label{Fig:3wn} (color online)
  The weight distributions of
   (a) the diagonal $w_n^{\rm diag}$,
   (b) the grand canonical $w_n^{\rm grand}$ and
   (c) the microcanonical $w_n^{\rm micro}$ density matrices as a
  function of energy $\omega =\alpha_\Lambda \Lambda^{-n/2}$, with
  $\alpha_\Lambda$ of order 1 (see text). These distributions were
  calculated after application of a single 1-particle excitation
  $\ket{\Psi} = \hat{f}_{k\up}^\dag\ket{\mathcal{G}}$ at shell $k$,
  where for a given Wilson shell $k$, the discretization parameter
  $\Lambda$ was chosen such that the excitation occurred at comparable
  energy, \ie $\Lambda(k) = E^{-2/k}$ for $E=T_K$. The vertical dashed
  lines in (a) present the corresponding energies at which the
  excitation occurred.
}
\end{figure}

\subsubsection{Single 1-particle excitation}

\Fig{Fig:3wn} shows the weights of the diagonal, grand canonical and
microcanonical density matrices calculated for a single 1-particle
excitation, $\ket{\Psi} = \hat{f}_{k\up}^\dag\ket{\mathcal{G}}$, for
different values of the discretization parameter $\Lambda$. The
discretization parameter was chosen such that the excitations
occurred at comparable energy, \ie $\Lambda (k) = E^{-2/k}$ for
fixed $E=T_K$. The weights are plotted as a function of energy
$\w=\alpha_\Lambda \Lambda^{-n/2}$, where $\alpha_\Lambda$ is a
discretization dependent but otherwise constant factor of order 1,
chosen such that the maximum of $w^{\rm grand}_n$ occurs at the same
energy for all $\Lambda$. The vertical dashed lines in
\fig{Fig:3wn}(a) show the corresponding energies $E_{\Psi}$, at
which the excitation was created, which without rescaling
by $\alpha_\Lambda$ would have
occurred at approximately the same energy. The weights of the diagonal
ensemble in \fig{Fig:3wn}(a) were averaged over two consecutive
Wilson shells to suppress the otherwise strong even-odd
oscillations, in particular, for larger $\Lambda$. In stark
contrast, $w_n^{\rm grand}$ in \fig{Fig:3wn}(b) does not exhibit
even-odd effects.\cite{Wb12tns}

By construction, the behavior of the weight for the microcanonical
ensemble in \fig{Fig:3wn}(c) is entirely different from $w_n^{\rm
diag}$ and $w_n^{\rm grand}$, in that $w_n^{\rm micro}$ is nonzero
only in a narrow energy window. Hence the following discussion will
focus on $w_n^{\rm diag}$ and $w_n^{\rm grand}$ in panels (a) and
(b), respectively. Also, as seen in \fig{Fig:3wn}(a), the decay of
the weight distribution at large energies, \ie for energies larger
than those at which the weight maximum occurs, is extremely fast for
both $w_n^{\rm diag}$ and $w_n^{\rm grand}$, much faster than
power-law or even exponential. In fact, by construction, $w_n^{\rm
diag}$ becomes strictly zero for Wilson shells larger than the shell
at which the excitation occurred.

In contrast, the decay for small energies, shows characteristic
powerlaw behavior. As seen in \fig{Fig:3wn}(a), $w_n^{\rm diag}$
decays linearly for small $\omega$, \ie $w_n^{\rm diag} \propto
\w^1$, which is thus independent of $\Lambda$. This is contrary to
$w_n^{\rm grand}$ [\fig{Fig:3wn}(b)], which decays
with $n$ like $d^{-n}$,\cite{WeichselbaumPRL07,Wb12tns}
for small energies (i.e. large $n$), where
$d=4$ is the dimension of the local state space of a single Wilson
site. Hence when plotted \vs $\omega$, $w_n^{\rm grand}$ becomes
dependent on the discretization parameter $\Lambda$. With the slope
for small $\omega$ in \fig{Fig:3wn}(b) decreasing with increasing
$\Lambda$, the slopes of $w_n^{\rm diag}$ and $w_n^{\rm grand}$
eventually coincide for large $\Lambda = d^2$ (having $d=4$ here,
$\Lambda=16$; not shown). Then, the diagonal and grand canonical
density matrices will have similar dependence on energy, and
therefore a single 1-particle excitation suffices to immediately
thermalize the system. However, with decreasing $\Lambda$, more and
more single-particle excitations will be required to fully
thermalize the system. Nevertheless, as will be shown below, for a
typical value of $\Lambda \geq 2$, still a very few 1-particle
excitations suffice to thermalize the system.

\begin{figure}[t]
\includegraphics[width=0.95\columnwidth,clip]{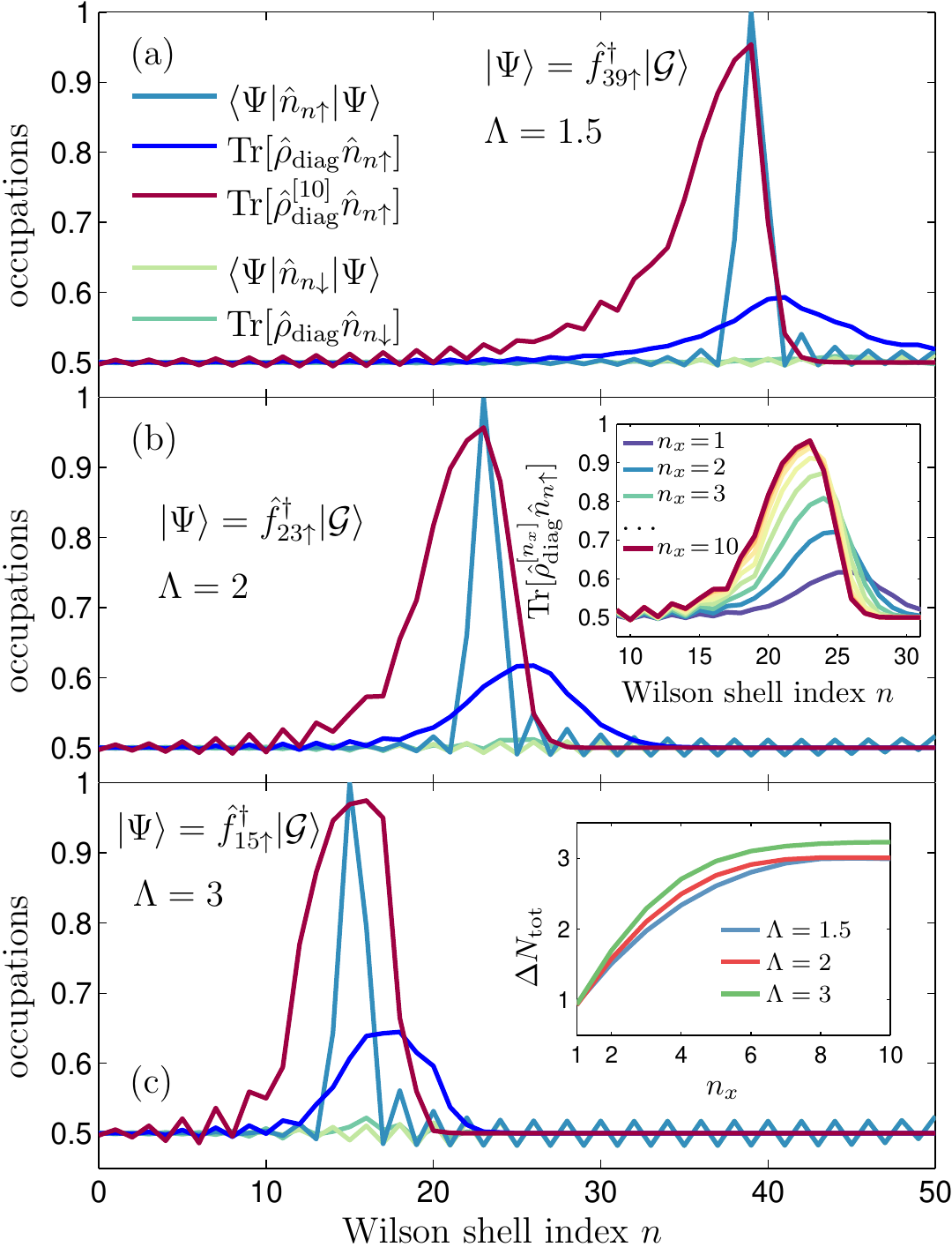}
  \caption{ \label{Fig:occup} (color online)
  The occupations on the Wilson chain for (a) $\Lambda=1.5$, (b)
  $\Lambda=2$ and (c) $\Lambda=3$ calculated for initial state (a)
  $\ket{\Psi} = \hat{f}_{39 \up}^\dag\ket{\mathcal{G}}$, (b)
  $\ket{\Psi} = \hat{f}_{23 \up}^\dag\ket{\mathcal{G}}$ and (c)
  $\ket{\Psi} = \hat{f}_{15 \up}^\dag\ket{\mathcal{G}}$ and for
  diagonal ensemble as a function of the Wilson shell index $n$. The
  excitation was created at the same energy scale for all $\Lambda$.
  The spin-up occupations obtained using $\hat{\rho}_{\rm diag}$ after
  having applied $n_x=10$ single-particle excitations are shown for
  comparison. The inset in (b) presents the buildup of spin-up charge
  \vs shell index $n$ when applying up to a total of $n_x$
  excitations. The inset in (c) shows the change of the total charge
  of the system $\Delta N_{\rm tot}$ as a function of $n_x$.
}
\end{figure}

\subsubsection{Multiple 1-particle excitations}

In \fig{Fig:occup} we show the occupations of the Wilson chain as a
function of shell index $n$, computed through $\bra{\Psi} \hat{n}_{
n\sigma}\ket{\Psi}$ and also by using the diagonal density matrix,
${\rm Tr} [\hat{\rho}_{\rm diag}\;\hat{n}_{n\sigma}]$, which
describes the effective long-time limit after a quench by a local
excitation. These occupations are calculated for three values of
discretization parameter $\Lambda \in \{1.5,\ 2,\ 3\}$ in panels
(a), (b), and (c), respectively.

We start by analyzing a state with a single excitation, \ie $n_x=1$
with $\ket{\Psi} = \hat{f}_{k \up}^\dag\ket{\mathcal{G}}$, where the
Wilson shell $k$ at which the operator acts, was chosen such that
the excitation energy was comparable ($\sim T_K$) for all $\Lambda$.
Thus, for $\Lambda=1.5$ [\fig{Fig:occup}(a)] the excitation was
created at shell $k=39$, for $\Lambda=2$ [\fig{Fig:occup}(b)] at
$k=23$, and for $\Lambda=3$ [\fig{Fig:occup}(c)] at $k=15$.
Irrespective of the value of discretization parameter, the behavior
of the chain occupations is qualitatively the same. First, since the
spin-up creation operator was applied, the spin-down occupations are
hardly affected. Second, $\bra{\Psi}\hat{n}_{n\uparrow}\ket{\Psi}$
reaches unity at the shell $k$ on which the excitation was created
and then for $n>k$, it shows somewhat more pronounced even-odd
oscillations. Third, after having evolved the system to infinite
time, the additional particle that had been inserted locally, smears
over several Wilson shells. However, the underlying logarithmic
discretization of the Wilson chain prohibits the particle to leave
the range of sites that represent its energy shell.

\begin{figure}[t]
\includegraphics[width=0.95\columnwidth,clip]{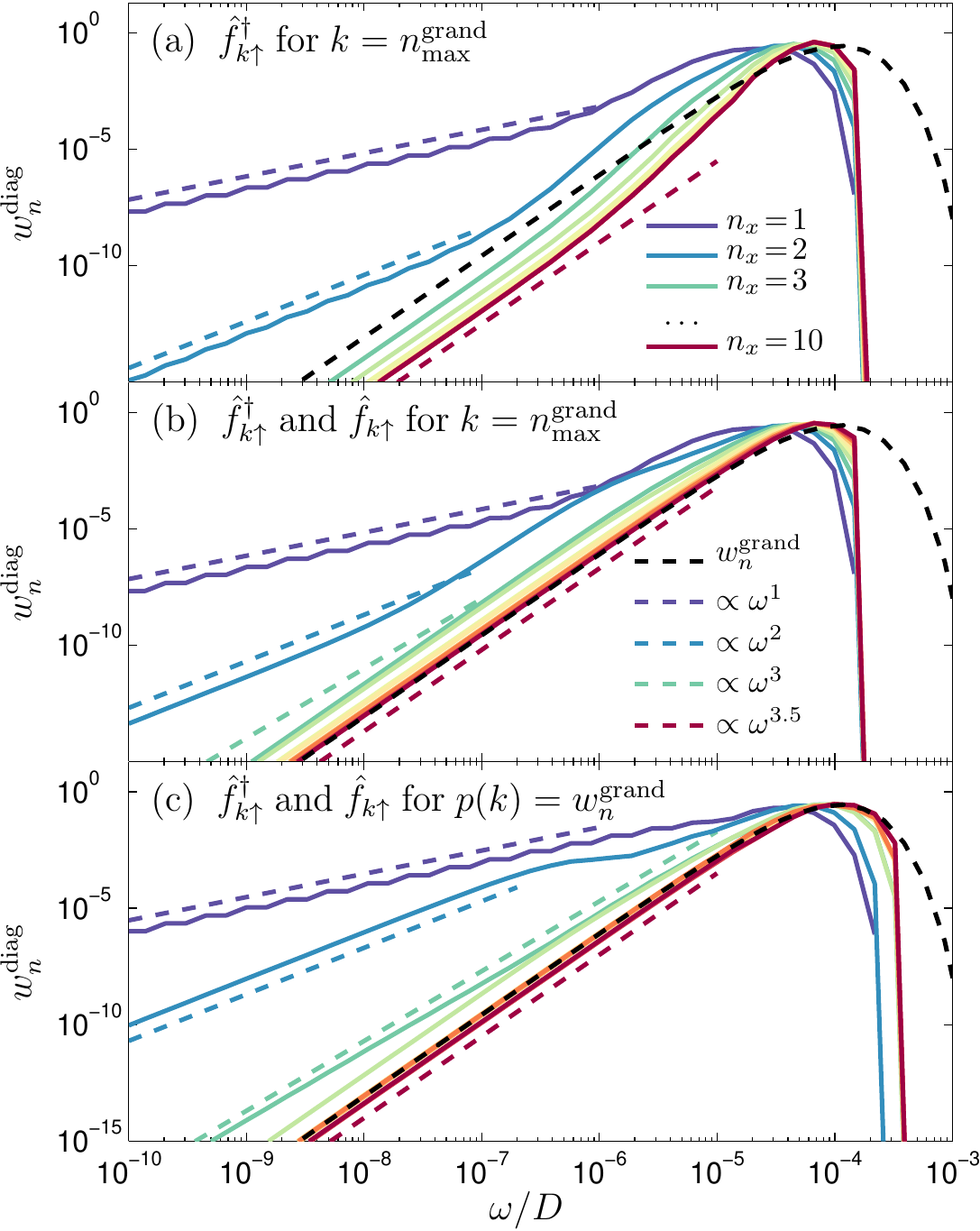}
  \caption{ \label{Fig:weights} (color online)
  The weights of the diagonal $w_n^{\rm diag}$ density matrix plotted
  on logarithmic scale as a function of energy $\w\propto
  \Lambda^{-n/2}$ calculated for single-particle excitations applied
  to the ground state ($n_x=1$) and then consecutively to the diagonal
  density matrix $(n_x>1)$. In (a) $\hat{f}_{k\uparrow}^\dag$ was
  applied at shell $n^{\rm grand}_{\rm max}$, at which $w_n^{\rm
  grand}$ is maximum. In (b) the operators $\hat{f}_{k\uparrow}^\dag$
  and $\hat{f}_{k\uparrow}$ were applied at shell $n^{\rm grand}_{\rm
  max}$ in an alternating way, whereas in (c) the same operators were
  applied with probability $p(k)$ determined by $w_n^{\rm grand}$. The
  dashed lines show the corresponding asymptotic lines. After
  $n_x\gtrsim 5$ excitations the slope of $w_n^{\rm diag}$ becomes
  equal to that of $w_n^{\rm grand}$ (black dashed line). This figure
  was calculated for $\Lambda=2.21$, for which the dependence of
  $w^{\rm grand}_n$ on energy is $w^{\rm grand}_n\propto
  \omega^{3.5}$.
}
\end{figure}

As discussed above, within the NRG, a single 1-particle excitation
is typically close but not exactly sufficient to fully thermalize
the system for typical values of the discretization parameter
$\Lambda \gtrsim 2$.
We therefore test if it is possible to thermalize the system by
sequential application of multiple 1-particle excitations with
intermediate relaxation. For this, we start with the system in its
ground state, to which a first 1-particle excitation is applied
($n_x=1$) at some shell $k$ that resembles the energy scale of a
fixed temperature $T=T_K$. After we allow the system to equilibrate,
\ie to time-evolve to $t\to\infty$, it is described by the
corresponding diagonal ensemble, \ie the diagonal density matrix
$\hat{\rho}^{[1]}_{\rm diag} \equiv \hat{\rho}_{\rm diag}$ and the
weights $w^{\rm diag}_{n} \equiv  w^{\rm diag}_{[1];n}$. Now given
this state, which is no longer described by a pure density matrix,
we apply a second 1-particle operator ($n_x=2$), and again let it
time-evolve to infinity. This generates a new diagonal ensemble
described by the density matrix $\hat{\rho}^{[2]}_{\rm diag}$ and
weights $w^{\rm diag}_{[2];n}$. This procedure can be repeated up to
a total of $n_x$ excitations applied to the system, which results in
the equilibrated diagonal ensemble $\hat{\rho}^{[n_x]}_{\rm diag}$,
represented by the weight distribution $w^{\rm diag}_{[n_x];n}$. The
underlying MPS diagrams that describe the required numerical
procedures are discussed in detail in \app{Ap:rhodiag}.

The occupations of the Wilson chain as a function of shell index $n$
for different number $n_x$ of 1-particle excitations applied to the
system are shown in the inset of \fig{Fig:occup}(b), while ${\rm Tr}
[\hat{\rho}^{[10]}_{\rm diag} \;  \hat{n}_{n\uparrow}]$ is
additionally shown in each panel of \fig{Fig:occup}. One can see
that repeated application of the same spin-up creation operator at
the same shell $k$ leads to an increase of the overall occupation
for spin-up. This increase, however, quickly saturates, as seen in
the analysis of the change of the total number of particles on the
Wilson chain $\Delta N_{\rm tot}$ in the inset to
\fig{Fig:occup}(c). For the rather broad range of discretizations
analyzed, $\Lambda \in [1.5,\, 3]$, repeated application of the same
creation operator with intermediate equilibration
allows to insert a total of at most three particles
into the system due to the underlying logarithmic
discretization. For the well-saturated value of $n_x=10$, the
inserted particles remain in the close vicinity of the shell that it
has been inserted.

A more detailed behavior of the system is provided by the weights of
the diagonal ensemble, which are shown in \fig{Fig:weights} as a
function of energy $\omega\propto\Lambda^{-n/2}$, for different
number $n_x$ of excitations created in the system. The case when the
same spin-up creation operator was applied to the shell at which the
grand canonical weights has its maximum, $k=n^{\rm grand}_{\rm
max}$, is presented in \fig{Fig:weights}(a). One can see that for
$n_x=1$ the low-energy tail of $w_n^{\rm diag}$ scales linearly with
energy [this exactly reflects the situation already seen in
\fig{Fig:3wn}(a)]. Now after applying the same operator a second
time ($n_x=2$), the low-energy dependence changes to $\propto
\omega^2$, and finally after $n_x\gtrsim 4$, the slope of $w_n^{\rm
diag}$ saturates and coincides with that of $w_n^{\rm grand}$ (at
ultrasmall $\omega$, $n_x=3$ still bends over to a $\omega^3$
powerlaw behavior; not shown). Similar behavior can be observed when
both creation and annihilation operators are applied in an
alternating way to the same shell $k=n^{\rm grand}_{\rm max}$, yet
as always with intermediate relaxation [see \fig{Fig:weights}(b)].
\Fig{Fig:weights}(c), finally, also relaxes the constraint of fixed
$k$, in that creation and annihilation operators are applied
alternatively at shell $k$ where $k$ is chosen with probability
$p(k)$ determined by the distribution of $w_n^{\rm grand}$
calculated at temperature $T_{\rm grand}=T_K$.
With this, the behavior of $w_n^{\rm diag}$ becomes even more
similar to that of $w_n^{\rm grand}$. The remaining sharp truncation
of the distribution towards large $\omega$ is due the finite
sampling in $n_x$. But as seen in \fig{Fig:weights}(c), this edge
can be moved towards larger frequencies (Wilson shells).

Note that the increasing exponent in the power law $\propto
\omega^{n_x}$ for the low energy behavior of the diagonal weight
distribution after $n_x$ applications of local creation or
annihilation operators with intermediate relaxation is perfectly
consistent with the concept of AO
as indicated in \eq{eq:AO:rho}:
with $\rho_{n;I=n_x}^\mathrm{diag} \propto \omega^{-n_x}$
and $\Delta n_\mathrm{loc}=1$, it follows
$w_{n;n_x+1}^\mathrm{diag} \propto \omega^{-(n_x+1)}$.
Specifically, at every iteration of $n_x$ another full particle needs to be moved
to or from infinity. For example, even if the number of particles in
the system saturates [inset to \fig{Fig:occup}(c)], application of
the same single-particle annihilation operator always meets with a
previously relaxed state and hence does not fully destroy the entire
state. However, (i) it projects out the part of the state where site
$k$ has been occupied, hence this is fully removed from the later
dynamics, and (ii) fills site $k$ for the (small) fraction of the
previously relaxed state which had no particle at site $k$.
Therefore in any case, as long as the applied operator does not
annihilate a given state such that the resulting state can again be
normalized, AO applies with integer exponents when creating or
annihilating particles without altering the Hamiltonian otherwise.
\cite{munder11}

The AO exponent that resembles the repeated quantum quenches
is only visible at the smallest energies, an energy scale that
quickly diminishes to zero with increasing $n_x$. Therefore
eventually the
exponent in the low-energy regime is limited by the NRG
exponent for a fully thermal distribution at a given finite
effective temperature $T$, i.e.
$w_n \propto d^{-n} = (\Lambda^{-n/2})^{2\log(d)/\log(\Lambda)}$
for $n > n_T$
[see \Sec{Sec:DM}]. This translates into a fully mixed state
space for $\omega\ll T$.

In \fig{Fig:weights}, the NRG discretization parameter
$\Lambda=2.21$ had been chosen such that the thermal distribution
$w_n^{\rm grand}$ has the non-integer exponent
$2\log(d)/\log(\Lambda) \simeq 3.5$ for the
low-energy tail. Nevertheless, repeated application of a few local
operators with integer AO exponents clearly allows to thermalize the
system to the corresponding
NRG specific thermal exponent. Within
the MPS framework, however, this just implies, that the remainder of
the system becomes completely randomized, since the grand-canonical
exponent simply implies full degeneracy with respect to the rest of the
system.

\begin{figure}[t]
\includegraphics[width=0.95\columnwidth,clip]{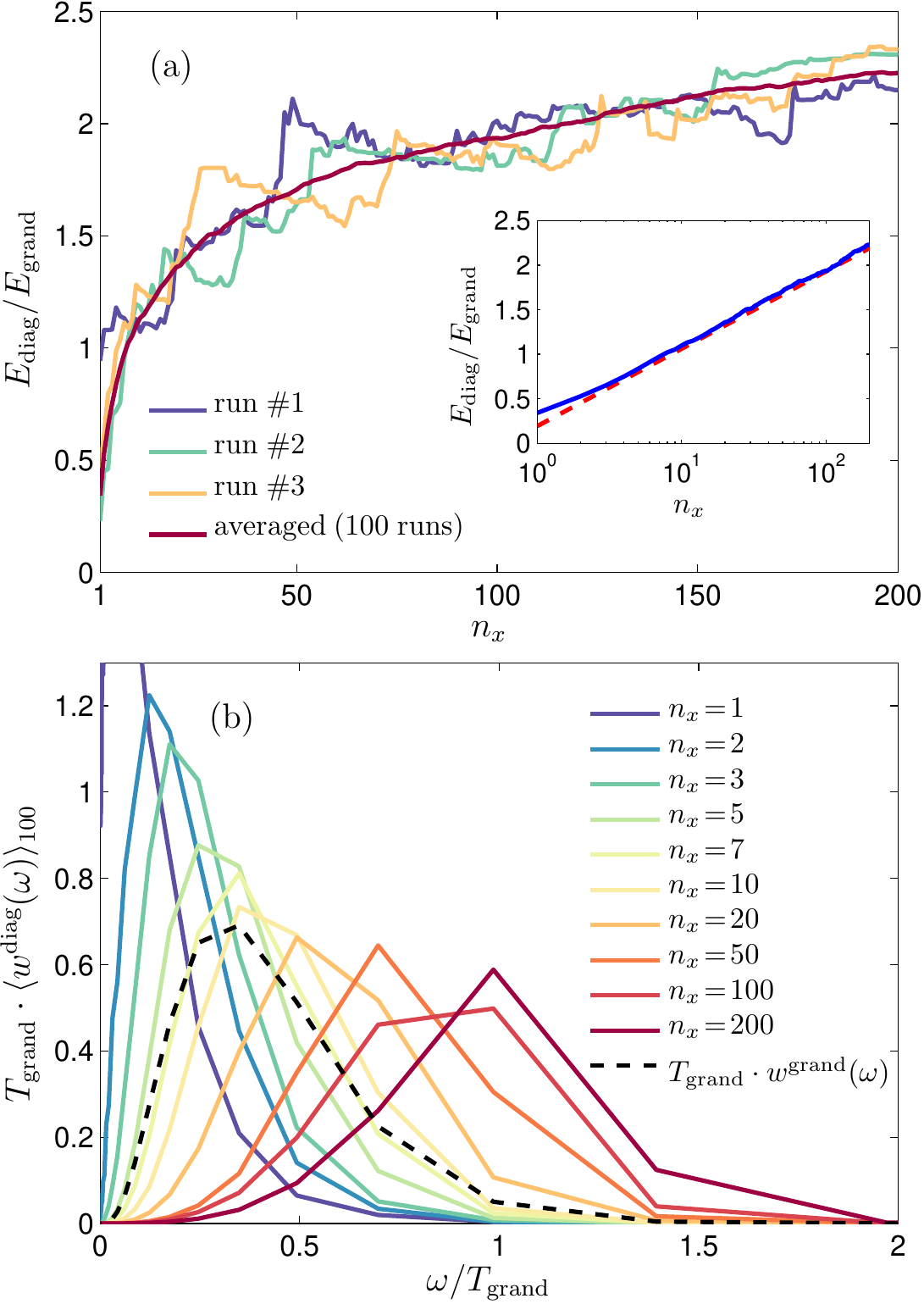}
  \caption{ \label{Fig:100avr} (color online)
  (a) Total energy $E_{\rm diag}$ of the system calculated using
  diagonal ensemble \vs the number of excitations $n_x$ and (b) the
  weights of the diagonal ensemble $w_n^{\rm diag}$ as a function of
  energy after having applied $n_x$ single-particle operators, either
  $\hat{f}_{k\uparrow}^\dag$ or $\hat{f}_{k\uparrow}$, selected
  randomly, and applied to shell $k$ with probability $p(k)$
  determined by $w^{\rm grand}_n$. $T_{\rm grand}$ is the temperature
  used for grand canonical ensemble and $E_{\rm grand}$ is the
  corresponding energy of the system. In (a) $E_{\rm diag}$ for three
  exemplary runs is shown together with its ensemble-averaged (over
  $100$ runs) value, which shows a steady logarithmic increase with
  $n_x$ [inset in (a)]. Panel (b) presents the ensemble-averaged
  weights $\langle w^{\rm diag}(\omega) \rangle_{100}$ (over $100$
  runs) of diagonal density matrix plotted as a function of energy
  $\omega$ with $w^{\rm diag}(\omega) \equiv w_{n}^{\rm
  diag}\cdot\Lambda^{n/2}$. The dashed line shows the weights of the
  grand canonical density matrix $w^{\rm grand}(\omega) \equiv
  w_{n}^{\rm grand}\cdot\Lambda^{n/2}$ as a function of $\omega$.
}
\end{figure}

Finally, an overall trend seen in all panels in \fig{Fig:weights} is
that by repeated applications of local quenches, the maximum of
$w_n^{\rm diag}$ shifts towards larger energies with $n_x$. Clearly,
this can be interpreted as heating of the system, and is analyzed in
detail by extending \fig{Fig:weights}(c) to significantly larger
$n_x$, with the results shown in \fig{Fig:100avr}. Panel (a)
analyzes the total energy $E_{\rm diag}$ of the system based on the
diagonal ensemble as a function of $n_x$ for three exemplary
individual stochastic runs by sampling a larger set of
excitations $\hat{f}_{k\sigma}^{(\dagger)}$ in $k$ and spin $\sigma$
according to a given temperature $T$ (see
caption of \fig{Fig:100avr}).
The energy $E_{\rm diag}$ is plotted in
units of $E_{\rm grand}$, \ie the total energy of the system
obtained from the grand canonical ensemble at temperature $T_{\rm
grand}=T_K$. We also show the ensemble average of $E_{\rm diag}$
over $100$ runs, which shows a slow but steady logarithmic increase,
as demonstrated by the inset in \fig{Fig:100avr}(a). This shows,
indeed, that the system is being heated up by the application of
multiple 1-particle excitations. Due to the underlying logarithmic
discretization, however, this heating is slowed down
logarithmically.

The ensemble-averaged weights $\langle w^{\rm diag} (\omega)
\rangle_{100}$ (averaged over $100$ runs) of diagonal density matrix
are shown in \fig{Fig:100avr}(b) as a function of energy $\omega$ on
a linear scale, where $w^{\rm diag}(\omega) \equiv w_{n}^{\rm
diag}\cdot\Lambda^{n/2}$ [the last factor is justified by the fact
that $w_{n}^{\rm diag}$ represents the integrated weight over an
energy interval of width $\propto \Lambda^{-n/2}$]. The dashed line
presents the weights of the grand canonical density matrix $w^{\rm
grand}(\omega) \equiv w_{n}^{\rm grand}\cdot\Lambda^{n/2}$. Again,
one can see that the maximum of $\langle w^{\rm diag}(\omega)
\rangle_{100}$ moves towards larger energies with $n_x$.

\section{Conclusions \label{sec:conclusions}}

In this paper we have studied the dynamics of the single-level
quantum dot in the Kondo regime, with a special focus on
thermalization. The dot was modeled by the single-impurity Anderson
Hamiltonian, which was analyzed by the numerical renormalization
group. Using the complete eigenbasis of the Hamiltonian obtained by
NRG, we calculated the spectral functions of the dot level by using
different statistical ensembles. In particular, we determined the
spectral function using the grand canonical, microcanonical and
diagonal ensembles, which were compared to the spectral function
calculated for a state, time-evolved with respect to the full
Hamiltonian. The main goal was to test whether the eigenstate
thermalization hypothesis also holds for truly many-body systems
displaying nontrivial correlations, such as the ones leading to the
Kondo effect. The ETH states that, in the long time limit, the
system can be described by relatively small number of representative
states, as given by the microcanonical ensemble.

We checked the validity of ETH for a few different states of the
system, including the ground state, the ground state of decoupled
dot, and a few states after having generated an excitation in the
bath with either single-particle or density operator. We showed that
for initial states where the excitation is created in the bath, the
microcanonical ensemble correctly describes the behavior of the
system in the long-time limit. Moreover, by calculating the
expectation values of the spectral function operator at the Fermi
level and the dot occupancy for energy eigenstates relevant for
microcanonical ensemble, we showed that the ETH can be indeed
invoked to understand the process of thermalization. The eigenstate
thermalization hypothesis is thus valid for initial states with an
excitation in the bath which, in the NRG context, can be interpreted
as macroscopic excitations. However, for states with an excitation
in the dot, the long time behavior cannot necessarily be described
by a statistical-mechanical ensemble and, as such, the reference to
ETH is not meaningful. This mandates an entirely different
interpretation in terms of quantum quenches.
In particular, when acting locally at the impurity, this essentially
leaves the system in its ground state, \ie at zero temperature.
Consequently, it clearly eludes the ETH, and therefore needs to be
distinguished from truly thermodynamic statistical ensembles.


\section*{Acknowledgments}

This work received support from the DFG (SFB-631, De-730/3-2,
De-730/4-2, SFB-TR12). IW acknowledges support from the Alexander
von Humboldt Foundation, the EU grant No. CIG-303 689 and the
National Science Centre in Poland as the Project No.
DEC-2013/10/E/ST3/00213, and AW in addition also DFG Grant
WE\-4819/1-1.

\appendix

\section{NRG implementation of different ensembles
\label{App:AdiagAmicro}}

We study the behavior of an expectation value of observable
$\mathcal{\hat{O}}$ in the long time limit under the time evolution
with respect to the Hamiltonian $\hat{H}$. The time evolution of a
many-body state of the system $\ket{\Psi}$ is given by,
$\ket{\Psi_t} = {\rm e}^{-i\hat{H}t} \ket{\Psi}$. Expanding in the
eigenbasis of the Hamiltonian, the time evolution of state
$\ket{\Psi}$ becomes,
\be
  \ket{\Psi_t} \cong
  \sum_{n}\sum_{se} {\rm e}^{-iE_{ns}^{D} t}\;
  C_{nse}^{D} \ket{se}_n^{D},
\ee
where the coefficients are defined as $C_{nse}^{D}\equiv {}^{D}_n
\bracket{se}{\Psi}$. Given an observable $\mathcal{\hat{O}}$, the
time evolution of its quantum mechanical expectation value, $O_t$,
is described by
\bea \label{Eq:expectO}
  O_t =  \sum_{nn'}\sum_{se\,s'e'}
  {\rm e}^{i\left( E_{ns}^{D}-E_{n's'}^{D} \right)t}\times \nonumber\\
  \left(C_{nse}^{D}\right)^*  \mathcal{O}_{nse,n's'e'}^{DD} \,
  C_{n's'e'}^{D} \,,
\eea
where $\mathcal{O}_{nse,n's'e'}^{DD} = {}^{D}_n
\bra{se}\mathcal{\hat{O}}\ket{s'e'}_{n'}^{D}$ denote the matrix
elements of the operator $\mathcal{\hat{O}}$. In general, the
evaluation of this expectation value is not trivial, as it includes
double sum over the discarded states of the Wilson chain. However,
as discussed in \app{Ap:nsum} [see Eq.~(\ref{Eq:2sum})], a double
sum over discarded states of the Wilson chain can be converted to a
single sum with contributions from both kept and discarded
states,~\cite{anders05a,anders05b} and the above formula becomes
\bea  \label{Eq:expectO2}
  O_t = \sum_{n}\sum_{ss'e}\sum_{XX'}^{\ne KK}
  {\rm e}^{i\left( E_{ns}^{X}-E_{ns'}^{X'} \right)t}  \times \nonumber\\
  \left(C_{nse}^{X}\right)^*  \mathcal{O}_{nse,ns'e}^{XX'}\, C_{ns'e}^{X'} \,.
\eea
In this way a time-dependent expectation value of any observable
$\mathcal{\hat{O}}$ can be calculated shell-wise in an iterative way
by performing a single sweep over the Wilson chain. Note that for
more complex operators $\mathcal{\hat{O}}$, such as the spectral
function operator $\hat{\mathcal{A}}$, the formula for the
expectation value, Eq.~(\ref{Eq:expectO}), may involve more sums
than just two, but it can still be written in a single-sum fashion,
with summations over different combinations of matrix elements
between kept and discarded states, except for the case when all
states are kept, see Eqs.~(\ref{Eq:3sum}) and (\ref{Eq:nsum}).

\subsubsection{Diagonal ensemble}

\begin{figure}[t]
  \includegraphics[width=0.9\columnwidth]{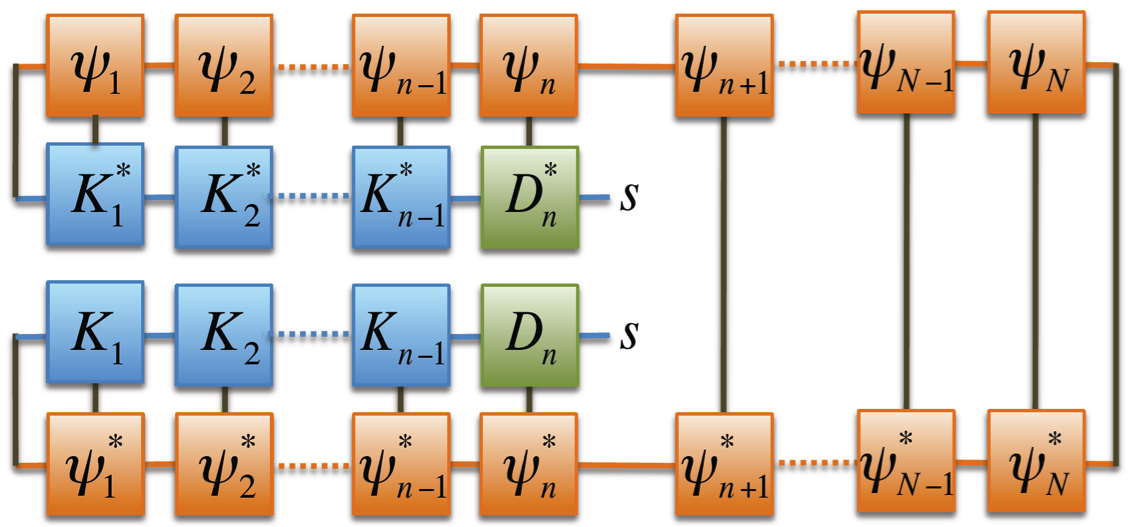}
  \caption{ \label{Fig:Cns} (color online)
  MPS representation of the coefficients
  $\sum_{e} |C_{nse}^{D}|^2 = \sum_{e}
  \bracket{\Psi}{se}^{D}_n  {}^{D}_n \bracket{se}{\Psi}$, where the
  discarded state $\ket{{se}}^{D}_n $ is defined graphically in
  Fig.~\ref{Fig:mps}. The kept and discarded blocks are denoted by
  $X=K$ ($X=D$), respectively, while the corresponding blocks of the
  state $\ket{\Psi}$ are denoted by $\psi_n$ for shell $n$. The star
  denotes complex conjugation.
}
\end{figure}

Equation~(\ref{Eq:expectO2}) implies that the infinite-time average
of the observable $\mathcal{\hat{O}}$ is given by a {\it diagonal}
ensemble\cite{Rigol_Nature08}
\be \label{Eq:Odiag}
  O_{\rm diag} =  \sum_{nse} \left|C_{nse}^{D}\right|^2
  \mathcal{O}_{nse,nse}^{DD} \;,
\ee
where the coefficient $\left|C_{nse}^{D}\right|^2$ gives the weight
of state $\ket{se}_n^D$ in the ensemble. These coefficients can be
used to define a {\it diagonal} density matrix,
\be
  \hat{\rho}_{\rm diag} \equiv
  \sum_{nse} \left|C_{nse}^{D}\right|^2 \;\ket{se}_n^{D}  {}_n^{D} \bra{se} \,.
\ee
It can be written in terms of shell-diagonal density matrices,
$\hat{\rho}_n^{\rm diag}$, as
\be
  \hat{\rho}_{\rm diag} \equiv
  \sum_{n} w^{\rm diag}_n \; \hat{\rho}^{\rm diag}_{n},
\ee
where $w_n^{\rm diag} \equiv \sum_{se} | C_{nse}^{D} |^2$, with
$\sum_n w_n^{\rm diag} = 1$, describes the cumulative weight of
shell $n$ to the density matrix $\hat{\rho}_{\rm diag}$. A diagram
for the calculation of the coefficients $\left|C_{nse}^{D}\right|^2$
is shown in Fig.~\ref{Fig:Cns}. By using Eq.~(\ref{Eq:Odiag}), the
expectation value of operator $\mathcal{\hat{O}}$ using the diagonal
density matrix can be written as
\be \label{Eq:Odiag2}
  O_{\rm diag} =  \sum_{n} w^{\rm diag}_n \;
  \trace_{s_n} (\hat{\rho}^{\rm diag}_{n} \; \mathcal{\hat{O}}_{n}) \;,
\ee
with $(\mathcal{O}_{n})_{ss'} 
\equiv {}_n^D\langle s|\mathcal{\hat{O}} | s' \rangle_n^D$,
where $\trace_{s_n}$ denotes the trace over states $s\in D$ at shell $n$.

\subsubsection{Microcanonical ensemble}

Now, according to the eigenstate thermalization
hypothesis,~\cite{Deutsch,Srednicki,Rigol_Nature08} to describe the
long time behavior of the system it is sufficient to consider just a
number of representative states of appropriate energy, as given by
the microcanonical ensemble:
\be \label{Eq:Omicro}
  O_{\rm micro}  = \frac{1}{N_{\Psi}}
  \sum_{\substack{nse \\ |E_{ns}^D - E_{\Psi}|\leq
  \delta E_{\Psi}}} \mathcal{O}_{nse,nse}^{DD}\;.
\ee
Here, $E_{\Psi}$ is the energy of state $\ket{\Psi}$ as in
\eq{eq:EPsi}, \ie relative to the ground state energy $E_0 =
\bra{\mathcal{G}}\hat{H}\ket{\mathcal{G}}$, with $\ket{\mathcal{G}}$
the full ground state of the system. Furthermore, $\delta E_\Psi$
characterizes the energy fluctuations, and $N_{\Psi}$ is the number
of energy eigenstates in interval $|E_{ns}^D - E_\Psi |\leq \delta
E_\Psi$.

\begin{figure}[t]
\includegraphics[width=0.99\columnwidth]{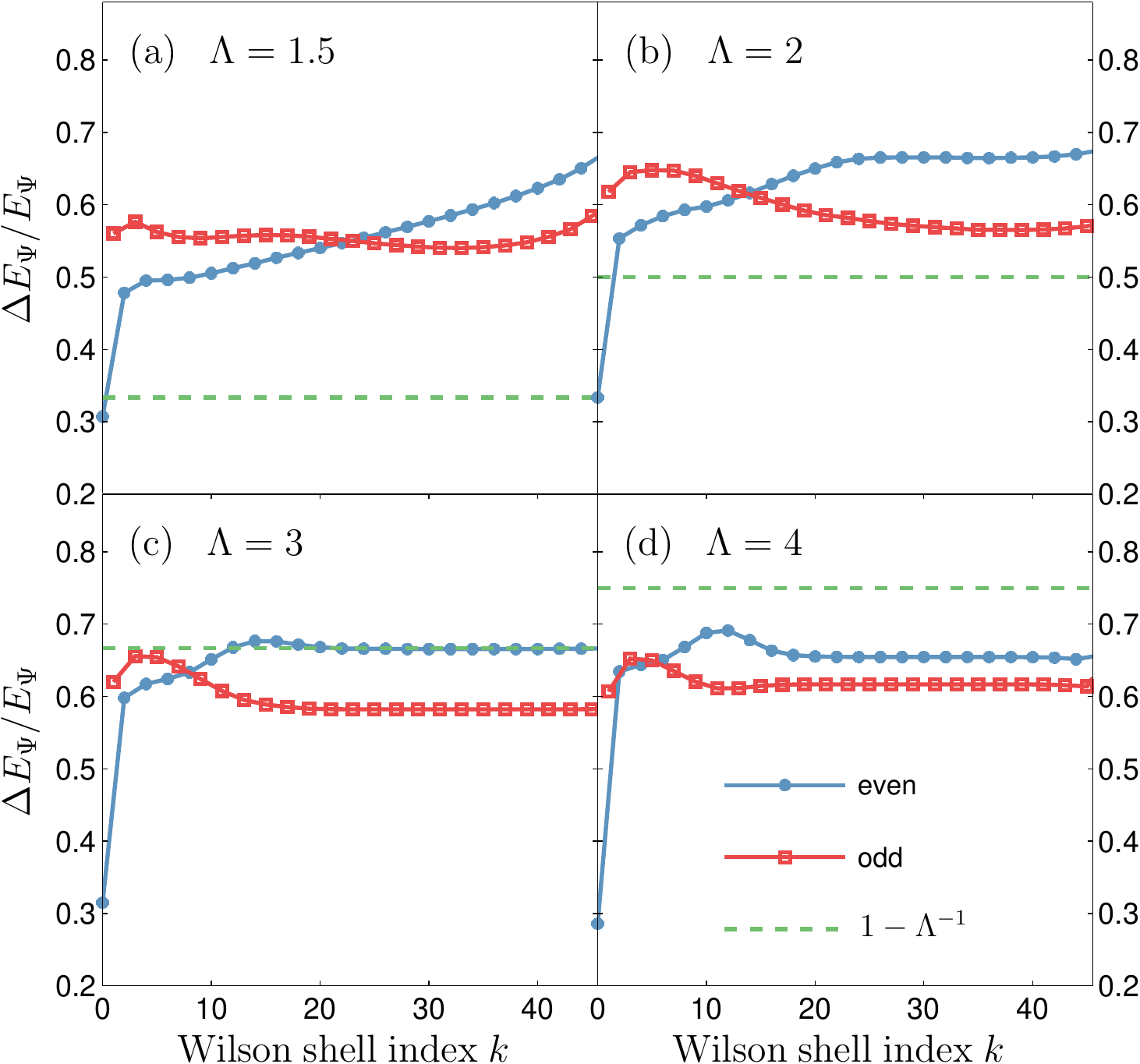}
  \caption{\label{Fig:DeltaE}
  (color online) The dependence of relative energy uncertainty
  $\Delta E_\Psi / E_\Psi$ on Wilson shell index for excited states
  $\ket{\Psi} = \hat{f}_{k\uparrow}^\dag \ket{\mathcal{G}}$
  and for different discretization parameters:
  (a) $\Lambda=1.5$, (b) $\Lambda=2$,
  (c) $\Lambda=3$, and (d) $\Lambda=4$.
  The horizontal dashed lines indicate the estimate for
  the relative energy uncertainty,
  $\delta E_\Psi / E_\Psi = 1 - \Lambda^{-1}$.
  }
\end{figure}

Regarding the energy uncertainty $\delta E_{\Psi}$, however,
consider first the relative energy uncertainty $\Delta E_\Psi /
E_\Psi$ as defined in \eq{eq:EPsi}. For excited states of the form
$\ket{\Psi} = \hat{f}_{k\uparrow}^\dag \ket{\mathcal{G}}$, this
is evaluated and presented in \fig{Fig:DeltaE} as a function of the
Wilson shell index $n$, where each panel corresponds to a different
discretization parameter $\Lambda$, as indicated. Clearly, this
needs to be distinguished from $\delta E_\Psi / E_\Psi$, which
determines the energy window that enters the microcanonical
expectation value $O_{\rm micro}$. While for an energy eigenstate
one trivially has $\Delta E_\Psi = 0$, for a general state
$\ket{\Psi}$, $\Delta E_\Psi / E_\Psi > 0$. As can be seen in
\fig{Fig:DeltaE}, for excited states obtained after acting with a
creation operator $\hat{f}_{k\uparrow}^\dag$ on the ground
state of the system, $\Delta E_\Psi / E_\Psi$ is smaller than unity,
but clearly finite, and for late iterations reaches $\Delta E_\Psi /
E_\Psi \approx 0.65$, irrespective of discretization parameter
$\Lambda$ [for panel (a) the Wilson shell index is not sufficiently
large to resolve the underlying low-energy fixed point, as seen in
the later panels; essentially the panels zoom out to reach smaller
energy scales for later panels, \ie with increasing $\Lambda$].
However, in the continuum limit, $\Lambda\to1^+$, one would expect
that $\ket{\Psi}$ in the long time limit behaves similar to an
energy eigenstate with energy $E_\Psi$ and the energy fluctuations
are suppressed, \ie $\delta E_\Psi \ll \Delta E_\Psi < E_\Psi$.
However, due to finite energy resolution when using NRG, it is not
possible to take $\delta E_\Psi \ll \Delta E_\Psi$, since otherwise
the energy window for the microcanonical ensemble in \eq{Omicro} may
contain only very few states or no states at all. Therefore, for the
results presented in the main paper, we took the finite energy
window for the microcanonical ensemble,
\begin{equation}
   \tfrac{\delta E_\Psi}{E_\Psi} =
   1-\tfrac{1}{\Lambda}
\text{,}\label{eq:dEmicro}
\end{equation}
which is also indicated by the horizontal dashed lines in
\fig{Fig:DeltaE}. This choice has the correct thermodynamic limit,
$\lim_{\Lambda\to 1^+} ( \delta E_\Psi / E_\Psi ) = 0^+$. The
motivation for \eq{eq:dEmicro} is given by the underlying
logarithmic discretization: if one considers the bath alone and
takes an arbitrary state at energy scale $E_\Psi$, then based on the
underlying single-particle energies, one expects an energy
resolution in terms of the single-particle level spacing $E_\Psi [
1-\Lambda^{-1}, 1+\Lambda ]$, the minimum of which was used to set
$\delta E_\Psi$. This choice guarantees that one has a comparable
number of representative energy eigenstates within a given energy
window. Moreover, with $\delta E_\Psi / E_\Psi$ clearly smaller than
$1$, only a finite window of Wilson shells of comparable width will
contribute to microcanonical ensemble for a given state
$\ket{\Psi}$.

For the microcanonical ensemble then one can use the complete NRG
eigenbasis to define a {\it microcanonical} density matrix,
$\hat{\rho}_{\rm micro}$, in the following manner. First, we
calculate the energy $E_{\Psi}$ for state $\ket{\Psi}$. Then, we
find all the eigenstates $\ket{se}_n^{D}$ of energy $E_{ns}^D$ that
obey, $|E_{ns}^D - E_{\Psi}|\leq \delta E_\Psi$, and the shells they
belong to, $n=n_1,\dots,n_2$. Let $N_n$ denote the number of such
states that contribute for given shell $n$. Then, the effective
total number of contributing states within the NRG is given by,
$N_{\Psi} = \sum_{n=n_1}^{n_2} N_n d^{n_2-n}$, where $d=4$ is the
dimension of the local space and we have also taken into account the
degeneracy of the environmental states. We can thus build a
normalized {\it mixed} density matrix for each shell,
$\hat{\rho}^{\rm micro}_{n}$. Knowing the mixed density matrices,
one can construct the full {\rm microcanonical} density matrix,
\be
  \hat{\rho}_{\rm micro} \equiv
  \sum_{n} w_{n}^{\rm micro}\; \hat{\rho}^{\rm micro}_{n},
\ee
where the weights $w_{n}^{\rm micro}$ take into account the effect
of degeneracies. They are given by $w_{n}^{\rm micro} = N_n
d^{n_2-n} / Z$, for $n=n_1,\dots,n_2$, and $w_n^{\rm micro} = 0$
otherwise, with the partition function $Z$ chosen such that $\sum_n
w_n^{\rm micro} = 1$. The microcanonical expectation value,
Eq.~(\ref{Eq:Omicro}), can be simply written then as,
\be 
O_{\rm micro}  =  \sum_{n} w^{\rm micro}_n \;
   \trace_{s_n}(\hat{\rho}^{\rm micro}_{n} \; \mathcal{\hat{O}}_{n})\;.
\ee

\subsubsection{Grand canonical ensemble}

The grand canonical expectation value $O_{\rm grand}$
can be obtained from
\be 
O_{\rm grand}  =  \sum_{n} w^{\rm grand}_n \;
   \trace_{s_n}(\hat{\rho}^{\rm grand}_{n} \; \mathcal{\hat{O}}_{n})\;,
\ee
with the density matrix and corresponding weights
as defined in Sec. \ref{Sec:DM}.

\section{Time-averaged spectral function
\label{Ap:At}}

\begin{figure}[t]
\includegraphics[width=0.9\columnwidth]{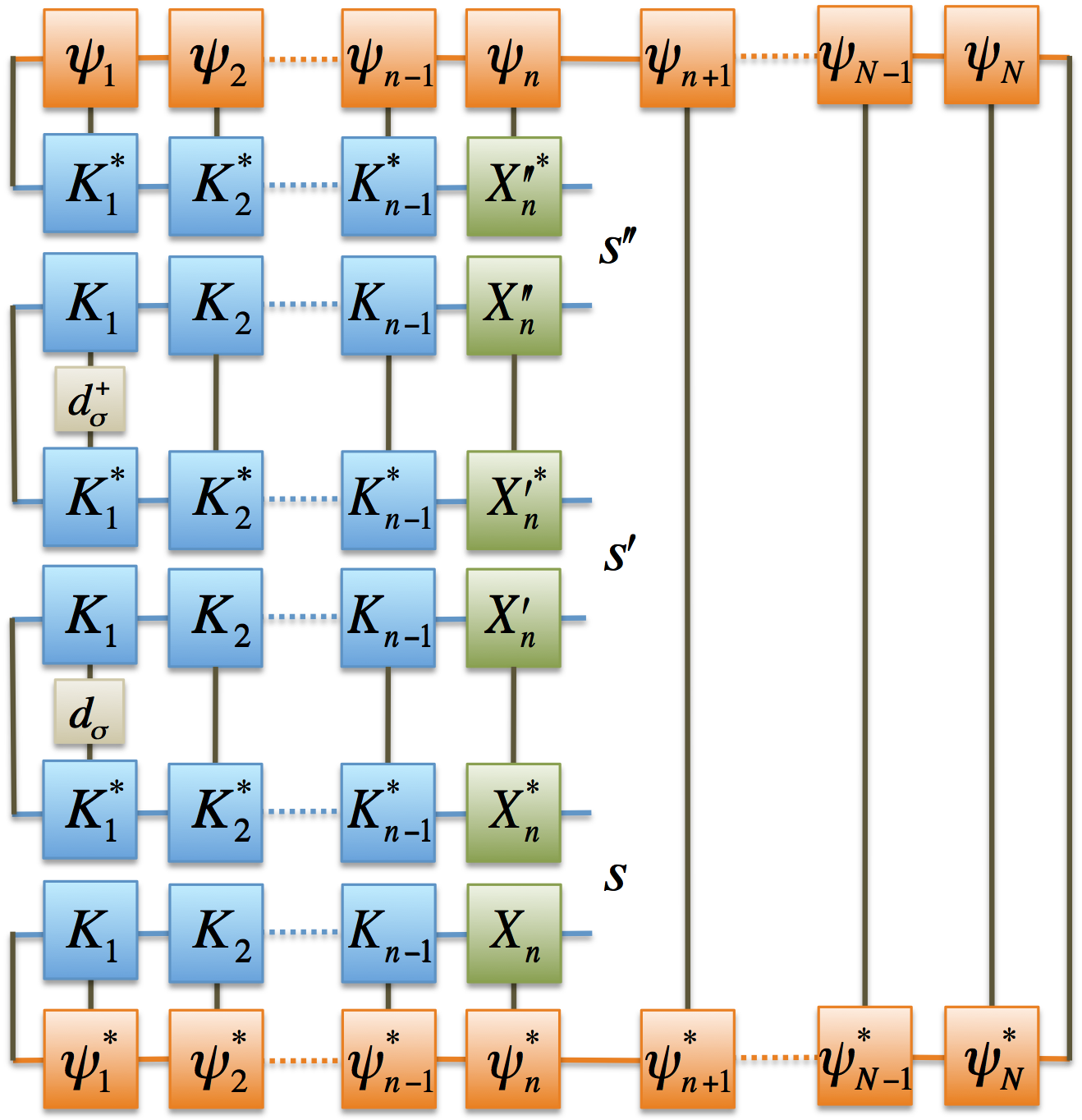}
\caption{ \label{Fig:MPS_Atime} (color online)
  MPS representation of a contribution to the time-averaged spectral
  function of the system due to $\hat{d}_\s(t)\hat{d}_\s^\dagger(0)$.
  The open indices need to be summed and multiplied with appropriate
  exponentials and matrix elements, see Eq.~(\ref{Eq:At2}). Summation
  over shells $n=\nzero,\dots,N$ and states $XX'X''\neq KKK$ also
  needs to be applied.
}
\end{figure}

The time-averaged spectral function can be calculated from
\eq{Eq:Atime}, where $A_{t}$ is explicitly given by [cf.
\eq{Eq:expectO2}]
\be \label{Eq:At1}
  A_{t} \!=\!\! \sum_{nss'e}\!\sum_{XX'}^{\ne KK} \!
  {\rm e}^{i\left( E_{ns}^{X}-E_{ns'}^{X'} \right)t} \!
  \left(C_{nse}^{X}\right)^{\!*} \!
  \mathcal{A}_{nse,ns'e}^{XX'}\; C_{ns'e}^{X'}\,.
\ee
The calculation is not trivial, since it in general involves triple
sums over discarded states of the Wilson chain. In particular, a
contribution to $A_t$ due to, e.g.,
$\hat{d}_\sigma^{}(\tau)\hat{d}_\sigma^\dag(0)$, is explicitly given
by
\bea \label{Eq:At2}
  A_{t}^{(1)} &=&
  \sum_{nss's''e}\sum_{XX'X''}^{\ne KKK}
  {\rm e}^{i\left( E_{ns}^{X}-E_{ns''}^{X''}
  \right)t} \, \delta(\w+E_{ns}^{X} - E_{ns'}^{X'})  \times \nonumber\\
  \lefteqn{ \left( C_{nse}^{X}\right)^* \, \left(d_\s \right)_{nse,ns'e}^{XX'}
                  \left(d_\s^\dagger \right)_{ns'e,ns''e}^{X'X''}
   C_{ns''e}^{X''} \;,
   }
\eea
the second line of which is illustrated in Fig.~\ref{Fig:MPS_Atime}.
Note that in \eq{Eq:At2} we have used the property (\ref{Eq:nsum})
to convert the triple sum over discarded states into a single sum
over the Wilson chain with contributions coming from all but $KKK$
states. Using \eq{Eq:At2}, the corresponding data points can be
calculated efficiently in a single-sweep fashion. At given iteration
$n$, one needs to perform all the contractions illustrated in
Fig.~\ref{Fig:MPS_Atime}, sum over the combinations of discarded and
kept states with $XX'X''\neq KKK$ while multiplying the open indices
with proper exponentials, and finally sum up the contributions from
all Wilson shells with $n = \nzero, \dots, N$. The second
contribution, $A_t^{(2)}$, to the time-averaged spectral function
$A_{\rm time}$ due to $\hat{d}_\sigma^\dag(0) \hat{d}_\sigma(\tau)$
has a form similar to \eq{Eq:At2}, with $\hat{d}_\s \leftrightarrow
\hat{d}_\s^\dagger$. The time averaging is eventually performed by
integrating the phase factors, $ {\rm exp}[i(E_{ns}^{X} -
E_{ns''}^{X''}) t]$, occurring in \eq{Eq:At2}, over time interval
$(t_{\rm fin}-\delta t,t_{\rm fin})$ and dividing by $\delta t$, see
\eq{Eq:Atime}. We also note that the calculation of $A_\Psi$, i.e.
for $\ket{\Psi_{t=0}}$, is much simpler than the calculation of
$A_{\rm time}$, since it requires only a double summation over
discarded states.

\section{Local operator applied to diagonal density matrix
\label{Ap:rhodiag}}

\begin{figure}[t]
  \includegraphics[width=1\columnwidth]{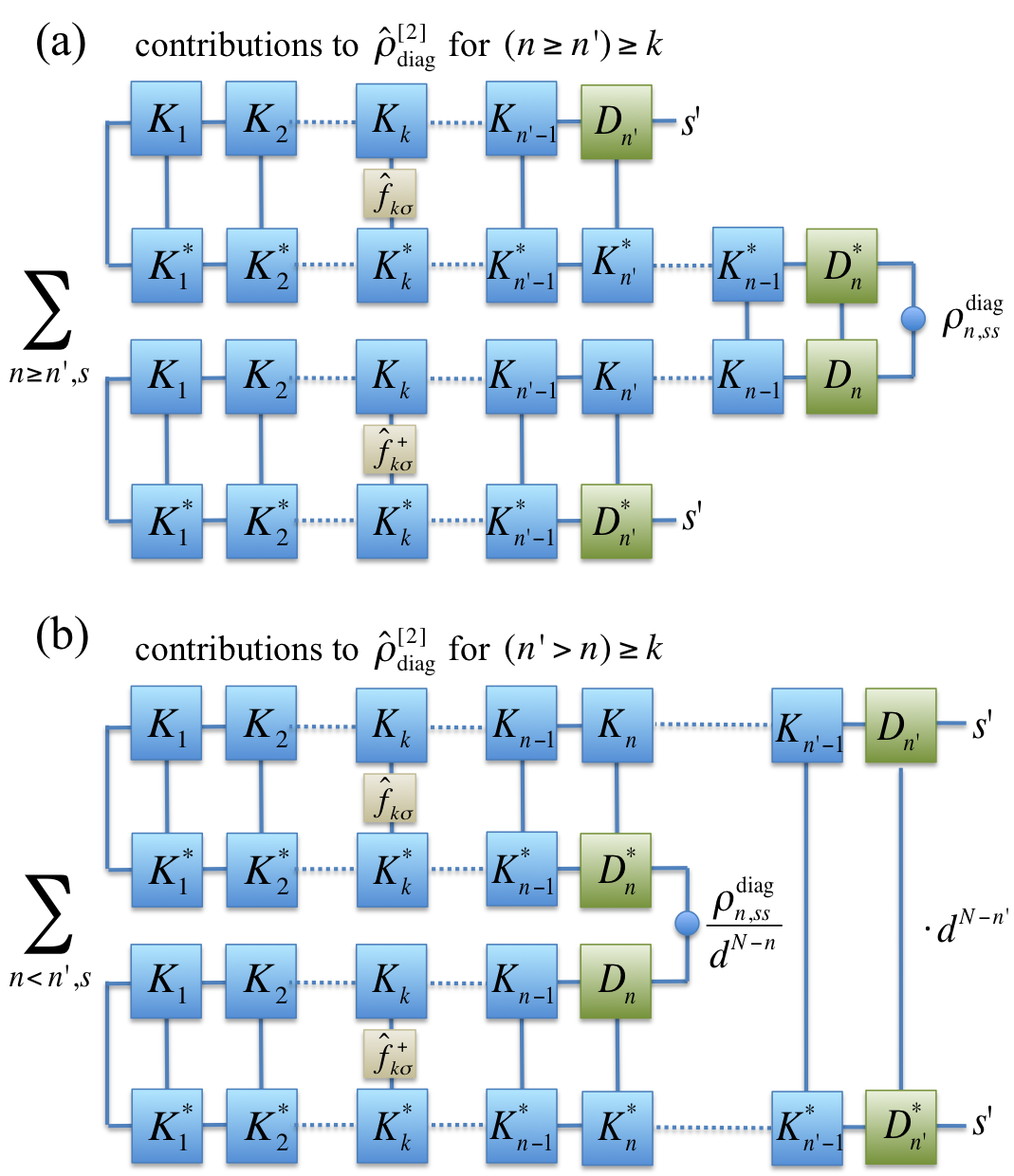}
  \caption{ \label{Fig:MPS_rhodiag} (color online)
  MPS representation of contributions to element of the new diagonal
  density matrix, $\rho_{n',s's'}^{\rm new}$, see
  Eqs.~(\ref{eq:rhodiagnew}) and (\ref{eq:rhodiagnew2}), after having
  applied a local operator, $\hat{f}_{k\sigma}^\dag$, at shell $k$.
  Note that consistent tracking of environmental degeneracy
  leads to the additional factor $d^{N-n'}/d^{N-n} = 1/d^{n-n'}$.
}
\end{figure}

Consider a density matrix $\hat{\rho}$. Its decomposition over the
Wilson shells $n$ will have the contributions $\hat{\rho}^{XX'}_n$,
with $XX'\neq KK$. In the long-time limit the off-diagonal matrix
elements average out and the density matrix becomes diagonal
\be
  \hat{\rho}_{\rm diag} = \sum_{ns}
  \ket{s}_{n}^{D} {}_{n}^{D} \bra{s} \cdot \rho_{n,ss}^{\rm diag} \,,
\ee
with $w_n^{\rm diag} = {\rm Tr} [\hat{\rho}_{n}^{\rm diag}]$, and
the environment \wrt shell $n$ already traced out and included in
$\rho_{n,ss}^{\rm diag}$. By construction of complete basis sets,
this density matrix is diagonal in the space of discarded states.
Reintroducing the state space for the full Wilson chain including
the environmental states $e$, it can be written as
\be \label{eq:rhodiag}
  \hat{\rho}_{\rm diag}^{[1]}
= \sum_{nse}   \ket{se}_{n}^{D} {}_{n}^{D} \bra{se} \cdot \frac{\rho_{n,ss}^{\rm diag}}{d^{N-n}} \,.
\ee
Now assume a 1-particle excitation is applied at shell $k$,
say $\hat{f}_{k\sigma}^\dag$, onto the given density matrix
$\hat{\rho}_{\rm diag}^{[1]}$. This results in
a new non-diagonal density matrix
\be
  \hat{\rho}_{\rm new} \equiv \hat{f}_{k\sigma}^\dag \, \hat{\rho}_{\rm diag} \, \hat{f}_{k\sigma}\,.
\ee
Again assuming that in the long-time limit the off-diagonal matrix
elements average out, this density matrix can be projected onto its
diagonal matrix elements, resulting in the new diagonal density
matrix
\be \label{eq:rhodiagnew}
  \hat{\rho}_{\rm diag}^{[2]} = \sum_{n's'e'}
  \ket{s'e'}_{n'}^{D} {}_{n'}^{D} \bra{s'e'} \cdot
  \rho_{n',s's'}^{\rm new} \;,
\ee
with the diagonal matrix elements
\begin{eqnarray}
  \rho_{n',s's'}^{\rm new}
&\equiv& \sum_{e'} 
  {}_{n'}^{D} \! \bra{s'e'}
     \hat{\rho}_{\rm new}
  \ket{s'e'}_{n'}^{D}
\label{eq:rhodiagnew2} \\
&=& \sum_{nse,e'} 
  {}_{n'}^{D} \! \bra{s'e'} \hat{f}_{k\sigma}^\dag \ket{se}_{n}^{D}
  \; \frac{\rho_{n,ss}^{\rm diag}}{d^{N-n}} \;
  {}_{n}^{D}\! \bra{se} \hat{f}_{k\sigma} \ket{s'e'}_{n'}^{D} \,.
\notag
\end{eqnarray}
This involves two independent summations over Wilson shells, which in the
usual spirit of Anders-Schiller (AS) basis \cite{anders05a,
anders05b} can be reduced to a single sum over shells based on
energy scale separation (see also \app{Ap:nsum}). Therefore the new
density matrix is defined by its contributions for each shell.
The main non-trivial contributions arise from the terms $(n,n')\ge k$, with
the contributions $n\ge n'$ ($n<n'$) depicted in
\fig{Fig:MPS_rhodiag}, panel (a) and (b), respectively, using the
MPS diagrams. Note that in panel (a) the degeneracy factor $d^{n-N}$
in $\hat{\rho}_{\rm diag}$, see Eq.~(\ref{eq:rhodiag}), again drops
out. In contrast, in panel (b) the overall degeneracy factor
acquires the correction $d^{n-n'}$.
The sum $\sum_{n}^{n\geq n'}$ in panel (a) can be computed
in a single prior backward sweep, similar to the full-density-matrix
(fdm)-NRG spirit.\cite{Wb12tns}
Moreover, also the sum $\sum_{n}^{k\leq n<n'}$ in panel (b) can be
computed in a single convoluted forward sweep.

In addition, there is one further relevant, yet simple contribution
to $\hat{\rho}_{\rm diag}^{[2]}$ where either $n$ or $n'$ are
smaller than $k$. The only non-zero
contribution of this type is $(n=n')<k$,
\be
  \rho_{n',s's'}^{\rm  new}
= \rho_{n',s's'}^{\rm diag} \cdot
  \expect{\hat{f}_{k\sigma}^\dag \hat{f}_{k\sigma}}_{T=\infty} =
  \frac{1}{2} \cdot \rho_{n',s's'}^{\rm diag} \,.
\ee
This, of course, is only relevant for iterations $n'$ where the
initial density matrix
had a finite contribution $w_{[1]; n'}^{\rm diag}$ to start with.

\section{Useful relations\label{Ap:nsum}}

In this appendix we collect some relations that are very useful for
explicitly performing (multiple) sums over complete sets of NRG
basis states. One such identity is that the sum over discarded
eigenstates of shells $n'>n$ is equal to the sum over kept
eigenstates of single shell $n$
\be \label{Eq:app1}
  \sum_{n'>n} \sum_{se}  \ket{se}_{n'}^{D} {}_{n'}^{D} \bra{se} =
  \sum_{se} \ket{se}_n^{K} {}_n^{K} \bra{se}.
\ee
Henceforth, this will be simply abbreviated by writing,
$\sum_{n'>n}^{D} = \sum_{n'=n}^{K}$. Note that the summation
$\sum_n^X$ is the shorthand notation for summing over discarded
($X=D$) or kept ($X=K$) states of a given shell $n$ and summing over
the shells. The above identity basically allows one to perform the
calculations of various expectation values in a single sweep over
the Wilson chain.

The double sum over the discarded states can be then written
as~\cite{anders05a,anders05b}
\bea \label{Eq:2sum}
  \sum_{nn'}^{DD} &=& \sum_{\substack{nn' \\ n=n'}}^{DD}
+ \sum_{\substack{nn' \\ n<n'}}^{DD} + \sum_{\substack{nn' \\ n>n'}}^{DD}
\nonumber\\
  &=&\sum_{n}^{DD} + \sum_{n}^{DK} + \sum_{n}^{KD} =
  \sum_{n}\sum_{XX'}^{\neq KK} \,,
\eea
where this double sum has been split into three terms, the first one
with $n=n'$ and the second (third) sum with $n<n'$ $(n>n')$, and
then the property (\ref{Eq:app1}) has been exploited.

In a similar way, for triple sum over discarded states we can then
write
\bea \label{Eq:3sum}
\sum_{nn'n''}^{DDD} &=&
\sum_{\substack{nn'n'' \\ n=n'=n''}}^{DDD}+
\sum_{\substack{nn'n'' \\ n<n'n''}}^{DDD}+
\sum_{\substack{nn'n'' \\ n'<nn''}}^{DDD}+
\sum_{\substack{nn'n'' \\ n''<nn'}}^{DDD}\nonumber\\
&&
+\sum_{\substack{nn'n'' \\ n=n'<n''}}^{DDD}
+\sum_{\substack{nn'n'' \\ n'=n''<n}}^{DDD}
+\sum_{\substack{nn'n'' \\ n=n''<n'}}^{DDD}
\nonumber\\
&=&
\sum_n^{DDD}+ \sum_n^{DKK} + \sum_n^{KDK} + \sum_n^{KKD}\nonumber\\
&&
+\sum_n^{DDK}+\sum_n^{KDD}+\sum_n^{DKD}\nonumber\\
&=&\sum_{n}\sum_{XX'X''}^{\neq KKK} \,.
\eea
Generalizing the above relations, expressions involving $M$ sums
over discarded states can be written in the following single-sum
form
\be \label{Eq:nsum}
  \sum_{n_1n_2\dots n_M}^{DD\dots D}
= \sum_n \sum_{X_1X_2\dots X_M}^{\neq KK\dots K} \,. \ee
This formula enables the calculation of various operator expectation
values in a single sweep fashion by collecting all the
kept-discarded contributions, excluding terms where all the states
belong to the kept state space.



\begin{thebibliography}{99}

\bibitem{sengupta04} 
K. Sengupta, S. Powell, and S. Sachdev, Phys. Rev. A {\bf 69}, 053616 (2004).

\bibitem{berges04} 
J. Berges, S. Borsanyi, and C. Wetterich, Phys. Rev. Lett. {\bf 93}, 142002 (2004).

\bibitem{krylov}
N. S. Krylov, {\it Works on the Foundation of Statistical Physics} (Princeton Univ. Press,
Princeton, 1979).

\bibitem{huang}
K. Huang, {\it Statistical Mechanics} (Wiley, New York, 1987).

\bibitem{Deutsch}
J. M. Deutsch, Phys. Rev. A {\bf 43}, 2046 (1991).

\bibitem{Srednicki}
M. Srednicki, Phys. Rev. E {\bf 50}, 888 (1994).

\bibitem{Rigol_Nature08}
M. Rigol, V. Dunjko, and M. Olshanii, Nature {\bf 452}, 854 (2008).

\bibitem{rigol09}
M. Rigol, Phys. Rev. Lett. {\bf 103}, 100403 (2009); Phys. Rev.
A {\bf 80}, 053607 (2009).

\bibitem{cassidy11} 
Amy C. Cassidy, Charles W. Clark, and Marcos Rigol,
Phys. Rev. Lett. {\bf 106}, 140405 (2011).

\bibitem{rigol12} 
Marcos Rigol and Mark Srednicki, Phys. Rev. Lett. {\bf 108}, 110601 (2012).

\bibitem{neuenhahn12} 
Clemens Neuenhahn and Florian Marquardt,
Phys. Rev. E {\bf 85}, 060101 (2012).

\bibitem{riera12} 
Arnau Riera, Christian Gogolin, and Jens Eisert, Phys. Rev. Lett. {\bf 108}, 080402 (2012).

\bibitem{ziraldo13} 
S. Ziraldo and G. E. Santoro, Phys. Rev. B {\bf 87}, 064201 (2013).

\bibitem{WilsonRMP75}
K. G. Wilson, Rev. Mod. Phys. {\bf 47}, 773 (1975).

\bibitem{BullaRMP08}
R. Bulla, T. A. Costi, and T. Pruschke, Rev. Mod. Phys. {\bf 80},
395 (2008).

\bibitem{Anderson}
P. W. Anderson, Phys. Rev. {\bf 124}, 41 (1961).

\bibitem{krishnamurthy80} 
H. R. Krishna-murthy, J. W. Wilkins, and K. G. Wilson, Phys. Rev. B {\bf 21}, 1003 (1980).

\bibitem{meir}
Y. Meir and N. S. Wingreen,
Phys. Rev. Lett. {\bf 68}, 2512 (1992).

\bibitem{meir93} 
Yigal Meir, Ned S. Wingreen, and Patrick A. Lee, Phys. Rev. Lett. {\bf 70}, 2601 (1993).

\bibitem{kondo64}
J. Kondo, Prog. Theor. Phys. {\bf 32}, 37 (1964).

\bibitem{hewson_book}
A. C. Hewson, {\it The Kondo Problem to Heavy Fermions} (Cambridge
University Press, Cambridge, 1993).

\bibitem{goldhaber-gordon_98}
D. Goldhaber-Gordon, H. Shtrikman, D. Mahalu, D. Abusch-Magder, U. Meirav, and M. A. Kastner,
Nature (London) {\bf 391}, 156 (1998).

\bibitem{cronenwett_98}
S. Cronenwett, T. H. Oosterkamp, and L. P. Kouwenhoven, Science
{\bf 281}, 182 (1998).

\bibitem{derWiel_Science00}
W. G. van der Wiel, S. De Franceschi, T. Fujisawa, J. M. Elzerman,
S. Tarucha and L. P. Kouwenhoven, Science {\bf 289}, 2105 (2000).

\bibitem{anders05a}
F. B. Anders and A. Schiller, Phys. Rev. Lett. {\bf 95}, 196801 (2005).

\bibitem{anders05b}
F. B. Anders and A. Schiller, Phys. Rev. B {\bf 74}, 245113 (2006).

\bibitem{WeichselbaumPRL07}
A. Weichselbaum and J. von Delft, Phys. Rev. Lett. {\bf 99},
076402 (2007).

\bibitem{petersPRB06}
R. Peters, T. Pruschke, and F. B. Anders, Phys. Rev. B {\bf 74}, 245114 (2006).

\bibitem{WeichselbaumPRB09}
A. Weichselbaum, F. Verstraete, U. Schollw\"ock, J. I. Cirac, and J. von Delft, Phys. Rev. B {\bf 80},
165117 (2009).

\bibitem{Haldane1978}
F. D. M. Haldane, Phys. Rev. Lett. {\bf 40}, 416 (1978).

\bibitem{rosch12}
A. Rosch, Eur. Phys. J. B {\bf 85}, 6 (2012).

\bibitem{anderson67}
P. W. Anderson, Phys. Rev. Lett. {\bf 18}, 1049 (1967);
Phys. Rev. {\bf 164}, 352 (1967);
K. D. Schotte and U. Schotte, Phys. Rev. 182, 479 (1969);
P. Nozieres, J. Gavoret, and B. Roulet, Phys. Rev. 178, 1084 (1969).

\bibitem{wb11aoc}
A. Weichselbaum, W. M\"under, and J. von Delft,
Phys. Rev. B 84, {\bf 075137} (2011).

\bibitem{munder11}
W. M\"under, A. Weichselbaum, M. Goldstein, Y. Gefen, and J. von Delft,
Phys. Rev. B {\bf 85}, 235104 (2012).

\bibitem{Wb12tns}
A. Weichselbaum, Phys. Rev. B {\bf 86}, 245124 (2012).

\end{thebibliography}
\end{document}